\documentclass[aps,prb,twocolumn,superscriptaddress,longbibliography]{revtex4-1}

\usepackage{graphicx}
\usepackage{dcolumn}
\usepackage{bm}
\usepackage{multirow}
\usepackage{amsthm}
\usepackage{amsmath}
\usepackage{amsfonts}
\usepackage{comment}
\usepackage{bbm}
\usepackage{graphicx}
\usepackage{subfigure}

\newtheorem{defn}{Definition}[section]

\newcommand{\Tr}{\mathrm{Tr}}
\newcommand{\bra}[1]{\mbox{$\langle #1 |$}}
\newcommand{\ket}[1]{\mbox{$| #1 \rangle$}}

\begin{document}

\author{Lexin Ding}
\affiliation{Faculty of Physics, Arnold Sommerfeld Centre for Theoretical Physics (ASC),\\Ludwig-Maximilians-Universit\"at M\"unchen, Theresienstr.~37, 80333 M\"unchen, Germany}

\author{Christian Schilling}
\affiliation{Faculty of Physics, Arnold Sommerfeld Centre for Theoretical Physics (ASC),\\Ludwig-Maximilians-Universit\"at M\"unchen, Theresienstr.~37, 80333 M\"unchen, Germany}
\affiliation{Wolfson College, University of Oxford, Linton Rd, Oxford OX2 6UD, United Kingdom}

\title{Correlation paradox of the dissociation limit: A quantum information perspective}

\date{\today}

\begin{abstract}
The interplay between electron interaction and geometry in a molecular system can lead to rather paradoxical situations.
The prime example is the dissociation limit of the hydrogen molecule: While a significant increase of the distance $r$
between the two nuclei marginalizes the electron-electron interaction, the exact ground state does, however, not take the
form of a single Slater determinant. By first reviewing and then employing concepts from quantum information theory, we resolve this paradox and its generalizations to more complex systems in a quantitative way. To be more specific, we illustrate and prove that thermal noise due to finite, possibly even just infinitesimally low, temperature $T$ will destroy the entanglement beyond a critical separation distance $r_{\mathrm{crit}}$($T$) entirely. Our analysis is comprehensive in the sense that we simultaneously discuss both
total correlation and entanglement in the particle picture as well as in the orbital/mode picture. Our
results reveal a conceptually new characterization of static and dynamical correlation in ground states by relating them
to the (non)robustness of correlation with respect to thermal noise.
\end{abstract}


\maketitle 


\section{Motivation}\label{sec:intro}
In recent years, the development of efficient descriptions of quantum many-body systems has been  strongly influenced by quantum information theory, particularly through the concept of entanglement.
From a general point of view, entanglement is one of the most fascinating concepts of modern physics. There are at least three reasons for its significance in various fields: (i) It provides important insights into the properties and behaviour of physical and chemical systems such as quantum phase transitions\cite{Osborne02,Vidal03,gu2004entanglement} and the formation/breaking of chemical bonds\cite{boguslawski2013orbital,Szalay17}, (ii) it serves as a diagnostic tool for the description of quantum many-body states in general \cite{boguslawski2012entanglement,szalay2015tensor}. Hence, its rigorous quantification facilitates the development of more efficient descriptions of strongly interacting systems \cite{Osborne02b,Vidal07}, (iii) it is an important resource used in the quantum information sciences for realizing, e.g., quantum cryptography\cite{Ekert91,Zeilinger00}, superdense coding\cite{bennett1992communication,fang2000experimental,ye2005scheme,schaetz2004quantum}  and possibly even quantum computing\cite{Jozsa03}.
In the more traditional fields such as condensed matter physics and quantum chemistry, however, point (iii) is not sufficiently acknowledged:
The quantification of entanglement and correlation in general is often flawed or at least operationally meaningless and the significance of the respective numbers for quantum information processing tasks is therefore unclear. Moreover, the relation between entanglement and total correlation is not sufficiently well understood.

An illustrative example for the questionable application of quantum information theoretical tools is the attempt to quantify the ``correlation'' contained in an $N$-electron quantum state in terms of the one-particle reduced density matrix, e.g., in Refs.~\onlinecite{Ziesche97b,huang2005entanglement}. The common reasoning is the following one: First, one defines the configuration states
\begin{equation}\label{Psiconf}
\ket{\Psi}=f_{\chi_1}^\dagger f_{\chi_2}^\dagger\cdot \ldots \cdot f_{\chi_N}^\dagger\ket{0}
\end{equation}
as being ``uncorrelated''. This seems to be plausible since ground states of \emph{non-interacting} electrons are exactly of that form
\eqref{Psiconf}, exhibiting a product structure of $N$ fermionic creation operators $f^\dagger_{\chi_j}$, populating the $N$ energetically lowest spin-orbitals $\ket{\chi_1},\ldots, \ket{\chi_N}$.

To apply the quantum information theoretical formalism which refers to \emph{distinguishable} subsystems one describes fermions by antisymmetric states within the Hilbert space $\mathcal{H}_1^{\otimes^N}$ of $N$ distinguishable particles  (``first quantization''). By referring to the tensor product $\mathcal{H}_1^{\otimes^N}$, each electron is assigned  \emph{its own} one-particle Hilbert space $\mathcal{H}_1$ and  algebra of observables and the notion of reduced density operators follows then accordingly.
Yet the unpleasant surprise is that even for an ``uncorrelated'' state \eqref{Psiconf} each of the $N$ electrons is still entangled with the complementary $N-1$ electrons. Indeed, the
von Neumann entropy
\begin{equation}\label{entropy}
S(\gamma)= -\mbox{Tr}[\gamma \log \gamma] = -\sum_{j} \lambda_j \log \lambda_j
\end{equation}
of the one-particle reduced density matrix (1RDM) $\gamma \equiv \Tr_{N-1}[\ket{\Psi}\!\bra{\Psi}]= 1/N\sum_{j=1}^{N}\ket{\chi_j}\!\bra{\chi_j}$ does not vanish.
One tries to ``fix'' this issue by normalizing $\gamma$ to the particle number $N$ instead. This has the effect that $\gamma$'s non-vanishing eigenvalues $\lambda_j$ change from $1/N$ to $1$ and $S$ would consequently vanish as desired \cite{Ziesche97}. Yet, the von Neumann entropy \eqref{entropy} has an information theoretical origin and meaning based on probability theory\cite{Jozsa97book,plenio2014introduction} which is now unfortunately lost.

In some systems of \emph{interacting} electrons the application of quantum information theoretical concepts is even more peculiar.
For instance, in the dissociation limit of the hydrogen molecule the ground state follows as
\begin{equation}\label{PsiH2inf}
\ket{\Psi}= \frac{1}{\sqrt{2}}\big(f_{L\uparrow}^\dagger f_{R\downarrow}^\dagger-f_{L\downarrow}^\dagger f_{R\uparrow}^\dagger\big)\ket{0}\,
\end{equation}
where $L,R$ denote the 1s orbital located at the left/right nucleus. The respective 1RDM normalized to $N=2$ follows as
\begin{equation}
\gamma = \frac{1}{2}\sum_{i=L/R}\sum_{\sigma=\uparrow/\downarrow}\ket{i\sigma}\!\bra{i\sigma}\,.
\end{equation}
This yields $S=2\log2 \neq 0$, despite the fact that both electrons are arbitrarily far separated which marginalizes their interaction. Driven by the latter fact, it has been suggested\cite{Ziesche97} to apply formula \eqref{entropy} in that specific case not to the full 1RDM but only to its orbital part, $\gamma_l = \sum_{i=L/R}\ket{i}\!\bra{i}$, obtained by integrating out the electron spin since this would yield $S=0$, as desired.

These (and many similar) examples for taking the profound meaning of quantum information theoretical concepts \emph{ad absurdum} have urged us to provide some clarification. In the form of the present paper we intend to contribute to the process of bringing together the needs of quantum information theorists (point (iii)) and quantum chemists and more traditional physicists (mainly interested in points (i) and (ii)). The dissociation limit of the hydrogen molecule will serve as an ideal example since it allows us to illustrate various relevant facets of the subject matter. At the same time, it is fundamentally relevant in quantum chemistry for illustrating the role of static correlations within the process of breaking chemical bonds: The ground state \eqref{PsiH2inf} in the dissociation limit cannot be written as a single configuration state (and is even not close to any) \emph{despite the fact} that the interaction between both electrons is marginalized due to their spatial separation. It is exactly this correlation paradox which elucidates the well-known difficulties of Kohn-Sham density functional theory in describing systems with multi-reference character\cite{Savin96,Cohen08}. In our work, we revisit this correlation paradox and provide a concise resolution of it.
To be more specific, we illustrate and prove that thermal noise due to finite, possibly even just infinitesimally low, temperature $T$ will destroy the entanglement beyond a critical separation distance $r_{\mathrm{crit}}$($T$) entirely. This rationalizes and clarifies the general (incorrect) perception that the ``correlation'' of the dissociated ground state vanishes. In that sense, from a general point of view, our work may add another facet to the electron correlation problem which has been one of the fundamental challenges in quantum chemistry since more than 50 years (see, e.g., Refs.~\onlinecite{Low58,Pop76,Bart78,Pop87,Mazz12}).

The paper is structured as follows. In Section \ref{sec:CandE} we review the quantum information theoretical concepts of correlation and entanglement between distinguishable subsystems and explain how to transfer them to the case of indistinguishable fermions. In Section \ref{sec:diss}, we introduce and diagonalize the Hubbard dimer model and resolve the correlation paradox in a qualitative way. This is followed in Section \ref{sec:resol} by a quantitatively concise resolution involving the correlation and entanglement measures introduced in Section \ref{sec:CandE}. A generalization of the correlation paradox and its resolution to more complex systems is provided in Section \ref{sec:general}. We conclude by offering a new characterization of static and dynamical correlation by referring to
the (non)robustness of the total correlation with respect to thermal noise.

\section{Concept of Correlation and Entanglement}\label{sec:CandE}

\subsection{The Quantum Information Theoretical Formalism}\label{sec:QIT}
The notion of correlation and entanglement plays a central role in quantum physics.
In this section, we review those concepts and their quantification in the common context of distinguishable subsystems as studied in quantum information theory. We restrict ourselves to the most important case of bipartite settings and refer the reader to Refs.~\onlinecite{Geza09,Horo09} for an introduction into the concept of multipartite correlation and entanglement.

To introduce the concepts of entanglement and correlation we first recall a few important aspects regarding quantum states and their geometry. Although it is illustrative to deal with wave functions, e.g.,\! $\psi(\Vec{r},\sigma)$, as one can use them to construct probability clouds for atomic and molecular orbitals visualization\cite{giessibl2001imaging, itatani2004tomographic}, it is advantageous to adopt the representation-free formalism of density operators, acting on an underlying (for simplicity finite-dimensional) Hilbert space $\mathcal{H}$. This facilitates more direct and compact definitions of correlation and entanglement. As a matter of fact, both concepts refer solely to a decomposition of the system into two (or more) subsystems and do not depend on any possible choice of basis states for those subsystems.
In this formalism, a quantum state is represented by a Hermitian operator $\rho$ that is positive (i.e., having non-negative eigenvalues) and trace-normalized to unity,  $\Tr[\rho]=1$, reflecting the probabilistic nature of quantum mechanics. In particular, the expectation value $\langle \hat{A} \rangle_\rho$ of an observable represented by a Hermitian operator $\hat{A}$
then follows according to Born's rule as
\begin{equation}
    \langle \hat{A} \rangle_\rho = \Tr[\rho \hat{A}].
\end{equation}
The quantum states $\rho$ form a convex (compact) space $\mathcal{D}$. Indeed,
any convex combination $\rho = p \rho_1 + (1-p)\rho_2$ of two density operators $\rho_1$ and $\rho_2$, $p \in [0,1]$, is again a density operator. The boundary of $\mathcal{D}$ is given by those $\rho \in \mathcal{D}$ which have at least one vanishing eigenvalue. In particular, the extreme points (those that cannot be written as a convex combination of others) are given by the pure states, $\rho \equiv \ket{\Psi}\!\bra{\Psi}$. From a general point of view, the space $\mathcal{D}$ of quantum states could be interpreted as a subset of the Hermitian matrices with $\mbox{dim}(\mathcal{H})$ many rows and columns. In that sense the space $\mathcal{D}$ can be equipped with a suitable metric. Examples include the distance metric based on the Frobenius norm, $d_F(\rho,\sigma)=\|\rho-\sigma\|_F\equiv\sqrt{\Tr[(\rho-\sigma)^2]}$, or the Bures distance
$d_B(\rho,\sigma)=\Tr[\sqrt{\sqrt{\rho}\sigma\sqrt{\rho}}]^2$. For further details on the geometry of density matrices and respective metrics we refer the reader to Refs.~\onlinecite{H92,PS96,B11}.

One of the important conclusions from those geometric considerations is that the possible similarity of two density operators $\rho, \sigma$ can be quantified in a universal way, i.e., without referring to a specific observable, despite the fact that $\rho,\sigma$ do not carry any physical unit. In particular, whenever two quantum states are close to each other in the state space $\mathcal{D}$, their expectation values will be close to each other for any observable as well. This follows directly from the Cauchy-Schwarz inequality, $|\langle \hat{A}, \hat{B}\rangle| \leq \sqrt{|\langle \hat{A}, \hat{A}\rangle|} \sqrt{|\langle \hat{B}, \hat{B}\rangle|}$ applied to the Hilbert-Schmidt inner product, $\langle \hat{A},\hat{B}\rangle \equiv \Tr[\hat{A}^\dagger \hat{B}]$,
\begin{eqnarray}\label{statesvsexp}
\big|\langle \hat{A} \rangle_{\rho}-\langle \hat{A} \rangle_{\sigma}\big| & = &\big| \Tr[\hat{A} (\rho-\sigma)] \big| \nonumber \\
&\leq & \|\hat{A}\|_F \,d_F(\rho,\sigma).
\end{eqnarray}
To fully appreciate relation \eqref{statesvsexp} let us recall that two quantum states with, e.g., the same energy can still differ in their expectation values of other relevant observables. A prominent example for this is the unrestricted Hartree-Fock ground state of the hydrogen molecule in the dissociation limit. It has the same energy as the exact ground state \eqref{PsiH2inf} but the incorrect spin expectation values.

All considerations so far were just referring to the total system.
The discussion of interesting physics refers, however, to a notion of subsystems. Let us consider in the following a quantum system which can be split into two subsystems $A$ and $B$. In the common quantum information theoretical formalism those two subsystems are assumed to be distinguishable and its states are described by density operators $\rho_{AB}$ on the total Hilbert space $\mathcal{H}_{AB}\equiv \mathcal{H}_A\otimes \mathcal{H}_B$, where $\mathcal{H}_{A/B}$ denotes the local Hilbert space of subsystem $A/B$. The underlying algebra $\mathcal{A}_{AB}$ of observables of the total system follows in the same way from the local algebras,
$\mathcal{A}_{AB}\equiv \mathcal{A}_{A}\otimes \mathcal{A}_{B}$. A particularly relevant class of observables
are the local ones, i.e, those of the form $\hat{A}\otimes \hat{B}$. As a matter of fact, they correspond to simultaneous measurements of $\hat{A}$ on subsystem $A$ and $\hat{B}$ on subsystem $B$. To understand the relation between both subsystems, one would be interested in understanding how the respective measurements of both local measurements are correlated. As a matter of definition, they are uncorrelated if the expectation value of $\hat{A}\otimes \hat{B}$ factorizes,
\begin{eqnarray}\label{ABzeroC}
\langle \hat{A}\otimes \hat{B}\rangle_{\rho_{AB}} &\equiv&\Tr_{AB}[\rho_{AB}\,\hat{A}\otimes \hat{B}] \nonumber \\
&=& \Tr_{AB}[\rho_{AB}\,\hat{A}\otimes \hat{1}_B]\, \Tr_{AB}[\rho_{AB}\,\hat{1}_A\otimes \hat{B}] \nonumber   \\
&\equiv & \Tr_{A}[\rho_{A}\,\hat{A}]\, \Tr_{B}[\rho_{B}\,\hat{B}] \equiv \langle \hat{A}\rangle_{\rho_{A}} \langle \hat{B}\rangle_{\rho_{B}}.
\end{eqnarray}
In the second line we introduced the identity operator $\hat{1}_{A/B} \in \mathcal{A}_{A/B}$ and the last line gives rise to the reduced density operators $\rho_{A/B}\equiv \Tr_{B/A}[\rho_{AB}]$ of subsystems $A/B$ obtained by tracing out the complementary subsystem $B/A$.
To quantify the correlation between the measurements of $\hat{A}$ and $\hat{B}$ one thus introduces the correlation function
\begin{equation}\label{ABcorfunc}
C_{\rho_{AB}}(\hat{A},\hat{B}) \equiv  \langle \hat{A}\otimes \hat{B}\rangle_{\rho_{AB}}-  \langle \hat{A}\rangle_{\rho_{A}} \langle \hat{B}\rangle_{\rho_{B}}.
\end{equation}
Popular examples are the spin-spin or the density-density correlation functions, i.e., the local operators $\hat{A}, \hat{B}$ are given by some spin-component operator $\hat{S}_\tau(\vec{x})$ or the particle density operator $\hat{n}(\vec{x})$ at two different positions $\vec{x}_{A/B}$ in space.

The vanishing of the correlation function for a specific pair of observables $\hat{A},\hat{B}$ does not imply by any means that the same will be the case for any other pair $\hat{A}',\hat{B}'$ of local observables. A prominent example would be again the dissociated hydrogen state \eqref{PsiH2inf}: Its electron density-density correlation function between the left ($L$) and right ($R$) side vanishes in contrast to the respective spin-spin correlation functions. Inspired by this example, one would like to introduce a measure for the correlation between both subsystems without referring to a specific pair of local observables. One idea would be to determine an average of the correlation function $C_{\rho_{AB}}(\hat{A},\hat{B})$  or its maximal possible value with respect to all possible choices of local observables $\hat{A},\hat{B}$. At first sight, those two possible measures of total correlation seem to be very difficult (if not impossible) to calculate for a given $\rho_{AB}$.
Yet, by referring to the geometric picture of density operators the introduction of a total correlation measure turns into a rather simple task. To explain this, we first define
\begin{defn}[Uncorrelated States]\label{def:uncorr}
Let $\mathcal{H}_{AB} \equiv \mathcal{H}_A \otimes \mathcal{H}_B$ be the Hilbert space and $\mathcal{A}_{AB}\equiv \mathcal{A}_A\otimes \mathcal{A}_B$ the algebra of observables of a bipartite system $AB$, with local Hilbert spaces $\mathcal{H}_{A/B}$ and local algebras $\mathcal{A}_{A/B}$. A state $\rho_{AB}$ on $\mathcal{H}_{AB}$ is called uncorrelated, if and only if
\begin{equation}
    \langle \hat{A} \otimes \hat{B} \rangle_{\rho_{AB}} = \langle \hat{A} \rangle_{\rho_A} \langle \hat{B} \rangle_{\rho_B},
\end{equation}
for all local observables $\hat{A}\in \mathcal{A}_A$, $\hat{B}\in \mathcal{A}_B$.
The set of uncorrelated states is denoted by $\mathcal{D}_0$ and states $\rho_{AB}\notin \mathcal{D}_0$ are said to be \textit{correlated}.
\end{defn}
A comment is in order regarding the local algebras $\mathcal{A}_{A/B}$ that playing a crucial role in Definition \ref{def:uncorr}. In the context of distinguishable subsystems one typically assumes that $\mathcal{A}_{A/B}$ comprises all Hermitian operators on the local space $\mathcal{H}_{A/B}$. As a consequence, a state $\rho_{AB}$ is then uncorrelated if and only if it is a product state, $\rho_{AB}= \rho_A \otimes \rho_B$. This conclusion is, however, not true anymore if one would consider in Definition \ref{def:uncorr} smaller subalgebras\cite{zanardi2001virtual}. Actually, exactly this will be necessary in fermionic quantum systems due to the number parity superselection rule \cite{SSR} (as discussed in Section \ref{sec:mode}).

\begin{figure}[htb]
    \centering
    \includegraphics[scale=0.23]{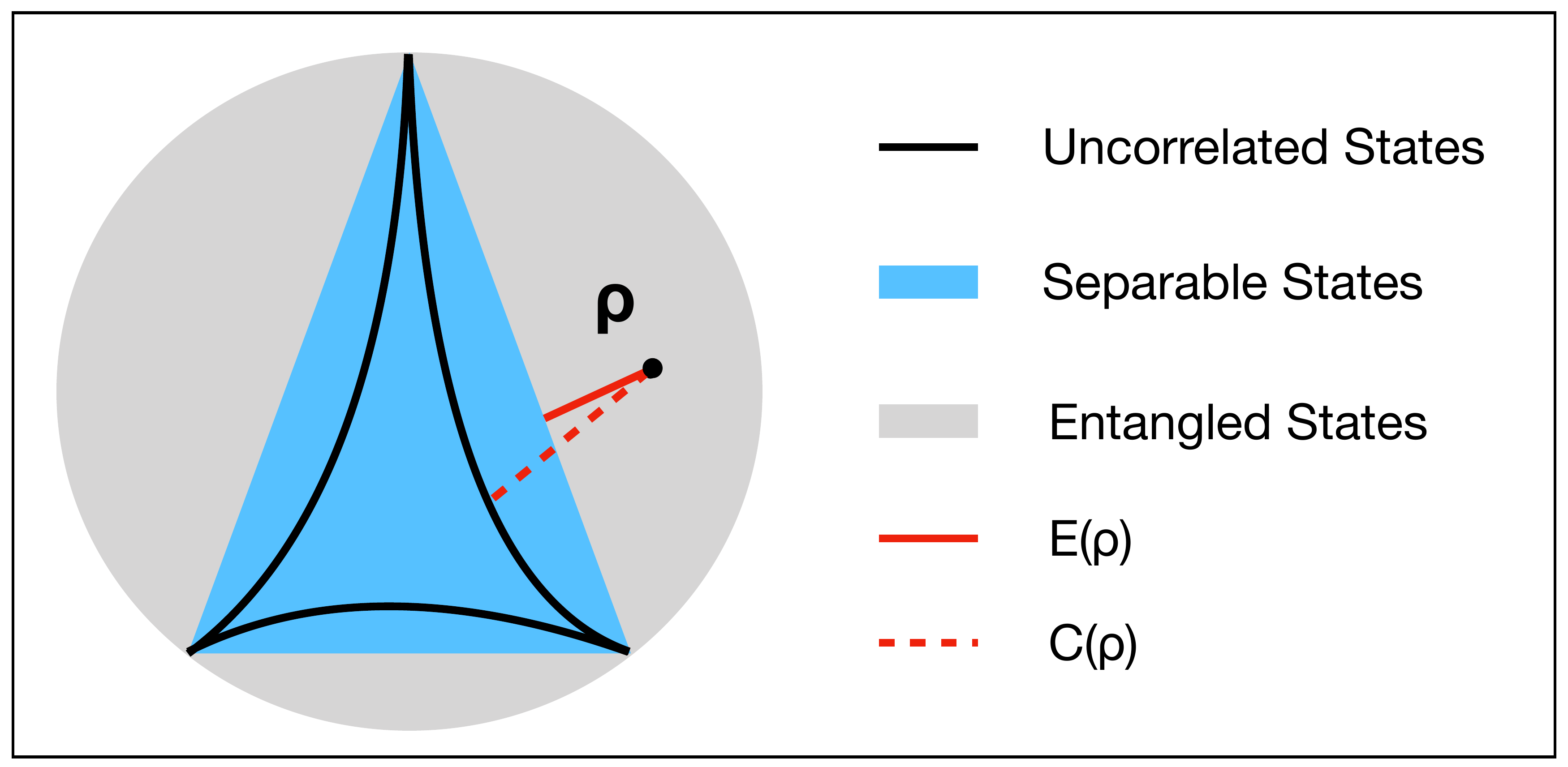}
    \caption{Schematic illustration of the space $\mathcal{D}$ of quantum states. Family $\mathcal{D}_0$ of uncorrelated states shown as thick black line and the separable states (its convex hull $\mathcal{D}_{sep}$) reside in the blue area. The grey area represents the entangled states and the red (dashed) line the geometric entanglement (correlation) measure $E(\rho)$ ($C(\rho)$).}
    \label{fig:states}
\end{figure}

By referring to the geometric picture of density operators a measure for the total correlation between $A$ and $B$ follows naturally. It is given by the distance of $\rho_{AB}$ to the set $\mathcal{D}_0$ of uncorrelated state (see also Figure \ref{fig:states} for an illustration). In principle one could base such a measure on any possible distance-function.
Yet, the notion of correlation and entanglement is formalized in quantum information theory by imposing plausible axioms defining valid measures, complemented by preferable features to guarantee an operational meaning \cite{Schum96,popescu1997thermodynamics,Geza09,Horo09}. While further details on that subject matter would go beyond the scope of the present work, we just would like to stress that the quantum relative entropy,
\begin{equation}\label{relEnt}
    S(\rho||\sigma) = \Tr[\rho(\log(\rho)-\log(\sigma))],
\end{equation}
emerges as the preferable underlying function for a geometric correlation (and entanglement) measure (despite the fact that it is not a distance function in the strict mathematical sense)\cite{vedral2002role}.
Besides its information theoretical meaning, the quantum relative entropy has additional appealing properties. For instance, it is invariant under unitary transformations,
\begin{equation}\label{Sunitary}
    S(\rho||\sigma) = S(U\rho U^\dagger || U \sigma U^\dagger)
\end{equation}
and it is convex in both arguments.
The total correlation measures follow as\cite{Lindblad73,vedral1997quantifying}
\begin{eqnarray}\label{CorrMeasure}
    C(\rho_{AB}) &\equiv& \min_{\sigma_{AB} \in \mathcal{D}_0} S(\rho_{AB}||\sigma_{AB}) \nonumber \\
    &=& S(\rho_{AB} || \rho_A \otimes \rho_B)\equiv I_{\rho_{AB}}.
\end{eqnarray}
Remarkably, the distinguished properties of the quantum relative entropy allow one to determine the minimizer $\sigma_{AB} \in \mathcal{D}_0$ of $\rho_{AB}$'s distance to $\mathcal{D}_0$ analytically. It follows as $\sigma_{AB}=\rho_A \otimes \rho_B$ and the correlation is nothing else than the quantum mutual information $I_{\rho_{AB}}$ (for the sake of completeness, we present the well-known proof of Eq.~\eqref{CorrMeasure} in Appendix \ref{app:MI}). The latter has a clear information theoretical meaning which emphasizes the significance of the total correlation measure \eqref{CorrMeasure}. It quantifies the information content in the state $\rho_{AB}$ which is not yet contained in the local states $\rho_A,\rho_B$.
We conclude by stating a crucial relation\cite{wolf2008area} between the total correlation \eqref{CorrMeasure} and individual correlation functions \eqref{ABcorfunc},
\begin{equation}\label{CvsI}
\frac{C_{\rho_{AB}}(\hat{A},\hat{B})}{\|\hat{A}\|_F\|\hat{B}\|_F} \leq \sqrt{2 \log(2)}\, \sqrt{I_{\rho_{AB}}}.
\end{equation}
This means in particular that the correlation function of \emph{any}  two local observables $\hat{A},\hat{B}$ is small whenever the quantum mutual information is small. The relation \eqref{CvsI} can be proven by combining \eqref{statesvsexp} applied to the observable $\hat{A}\otimes \hat{B}$ with a well-known relation between the quantum relative entropy and the Frobenius norm (see, e.g., Theorem 10.6 in Ref.~\onlinecite{Watrous}).

A possibly large total correlation suggests that
the accurate description of the total system $AB$ requires significantly more computational effort than the one of both individual subsystems $A,B$ (in case they were entirely decoupled).  While this is rather unfortunate for a quantum chemist or a condensed matter theorist (they are interested in an accurate description of such systems) the opposite is true from a quantum informational point of view. To be more specific, primarily the quantum part of the total correlation represents an important resource for realizing quantum information processing tasks such as  quantum cryptography\cite{Ekert91,Zeilinger00}, superdense coding\cite{bennett1992communication,fang2000experimental,ye2005scheme,schaetz2004quantum}, quantum teleportation \cite{bennett1993teleporting, bouwmeester1997experimental, furusawa1998unconditional}  and possibly even quantum computing\cite{Jozsa03}. Typical protocols for realizing such fascinating tasks utilize so-called Bell pairs, i.e., maximally entangled pure state of two qubits (two-level quantum systems), e.g.,
\begin{equation}
    |\Psi\rangle = \frac{1}{\sqrt{2}} \big( |\!\uparrow \rangle_A \otimes | \!\downarrow \rangle_B - |\!\downarrow \rangle_A \otimes |\!\uparrow\rangle_B \big). \label{Bell}
\end{equation}
It is thus one of the most important challenges to rigorously quantify the number of such Bell pairs that could be extracted from a given correlated quantum state $\rho_{AB}$.
It is not hard to imagine that a correlated quantum state $\rho_{AB}=\sum_i p_i \rho_A^{(i)}\otimes \rho_B^{(i)}$ which is given as the \emph{classical mixture} of uncorrelated states $\rho_A^{(i)}\otimes \rho_B^{(i)}$ does not offer any useful resource in that context: The system $AB$ is found in uncorrelated states $\rho_A^{(i)}\otimes \rho_B^{(i)}$ yet there is a classical probabilistic uncertainty in which of them it will be in. To elaborate further on the quantification of entanglement one defines
\begin{defn}[Separable States]\label{def:sep}
A state $\rho_{AB}$ is called separable/non-entangled if $\rho_{AB}$ can be expressed as a convex linear combination of uncorrelated states, that is $\rho_{AB} \in \mathrm{Conv}(\mathcal{D}_0) \equiv \mathcal{D}_{sep}$. Otherwise a state is called entangled.
\end{defn}
\noindent Here, $\mbox{Conv}(\cdot)$ stands for the convex hull and the term \textit{separable} is frequently used in quantum information theory for denoting non-entangled states. In this paper, we will use these two terms interchangeably.

In complete analogy to the concept of total correlation one has formulated plausible axioms for valid and operationally meaningful entanglement measures\cite{Geza09,Horo09}. By referring the reader to Ref.~\onlinecite{vedral1997quantifying} for more details, it is again the geometric picture which leads to a prominent entanglement measure, the \emph{relative entropy of entanglement}
\begin{equation}\label{EntMeasure}
    E(\rho_{AB}) \equiv \min_{\sigma_{AB} \in \mathcal{D}_{sep}} S(\rho_{AB}||\sigma_{AB}).
\end{equation}
It measures in terms of the quantum relative entropy \eqref{relEnt} the minimal distance of $\rho_{AB}$ to the set $\mathcal{D}_{sep}$ of separable states.
This and the general geometric picture is illustrated in Figure \ref{fig:states}. The set $\mathcal{D}_0$ of uncorrelated states (recall  Definition \ref{def:uncorr}) is shown as a black curve. According to Definition \ref{def:sep}, its convex hull $\mathcal{D}_{sep}\equiv \mbox{Conv}(\mathcal{D}_0)$ comprises all seperable/non-entangled states while the remaining density operators (gray area) are entangled. The geometric correlation and entanglement measures are given by the closest distance from a general state $\rho\equiv \rho_{AB}$ to the sets $\mathcal{D}_0$ (red dashed) and $\mathcal{D}_{sep}$ (red), respectively. Since the uncorrelated states are in particular non-entangled, $\mathcal{D}_0 \subset \mathcal{D}_{sep}$, the entanglement can never exceed the total correlation,
\begin{equation}\label{CorrvsE}
    C(\rho_{AB}) \geq E(\rho_{AB}).
\end{equation}

In contrast to the total correlation, there is no explicit analytical expression known for the relative entropy of entanglement in case of general mixed states and even its numerical calculation is typically quite demanding. That is quite different in case of pure states, $\rho_{AB}=\ket{\Psi_{AB}}\!\bra{\Psi_{AB}}$, since \eqref{EntMeasure} then simplifies to the entanglement entropy which is defined as the von Neumann entropy of the reduced density operators of subsystem $A$ and $B$, respectively \cite{popescu1997thermodynamics, vidal2000entanglement},
\begin{equation}\label{entanglepure}
    E(\ket{\Psi_{AB}}\!\bra{\Psi_{AB}}) = S(\rho_{A/B}) = -\Tr[\rho_{A/B} \log(\rho_{A/B})].
\end{equation}
It is equivalent to calculate the entanglement entropy with either $\rho_A$ or $\rho_B$, as they have the same eigenvalues in case of pure total states\cite{nielsen2002quantum} and one has $E(\ket{\Psi_{AB}}\!\bra{\Psi_{AB}}) = 0$ if and only if $|\Psi_{AB}\rangle$ factorizes.
In that context, we also would like to recall that for mixed states $\rho_{AB}$ the entanglement entropy \eqref{entanglepure} is obviously not a good measure for entanglement anymore since the mixedness of the reduced density operators $\rho_{A/B}$ could originate just from possible mixedness in $\rho_{AB}$ (classical correlation).

Finally, we would like to stress that for pure total states a remarkable operational meaning of the entanglement entropy (to the base $2$) has been found\cite{Schum96,popescu1997thermodynamics}: In the asymptotic limit of $n$ identical two-qubit systems, each in the same pure quantum state shared between two parties $A$ and $B$ with an entanglement entropy $S$, the number $m$ of maximally entangled Bell pairs that can be extracted follows as $m=  n\, S$. It is exactly this operational meaning of entanglement between distinguishable subsystems which raises some doubts about the common approach to entanglement and correlation in condensed matter physics and quantum chemistry: Applying some partial trace-like map to obtain some type of reduced density operator, possibly even not normalized to unity, and then plugging it into the formula for the von Neumann entropy does not necessarily mean to quantify correlation or entanglement.

\subsection{Fermionic Correlation}\label{sec:Ferm}
The concepts of entanglement and correlation, as reviewed in the previous section, refer to a well-defined separation of the total system into two (or more) distinguishable subsystems. In the simplest case, this separation emerges naturally from the physical structure of the total system, namely by referring to a possible spatial separation of two subsystems. In that case, it will be also easier to experimentally access both subsystems to eventually extract the entanglement from their joint quantum state. Nonetheless, the notion of bipartite correlation and entanglement is by no means unique for a given system since one just needs to identify some tensor product structure in the total system's Hilbert space, $\mathcal{H}\equiv \mathcal{H}_A\otimes \mathcal{H}_B$. In the most general approach, one even defines subsystems by choosing two commuting subalgebras $\mathcal{A}_A, \mathcal{A}_B$ of observables\cite{zanardi2001virtual}. This also highlights the crucial fact that entanglement and correlation are relative concepts since they refer to a choice of subsystems/subalgebras of observables.

In case of identical fermions the identification of subsystems is not obvious at all. For instance, how could one decompose the underlying $N$-fermion Hilbert space $\mathcal{H}_N \equiv \wedge^N[\mathcal{H}_1]$ or the Fock space $\mathcal{F}\equiv \oplus_{N\geq 0}\mathcal{H}_N$?
Actually, there exist two natural routes which look promising. The first one (presented in Section \ref{sec:mode}) refers naturally to the 2nd quantized formalism and leads to a notion of \emph{mode} (sometimes also called orbital or site) entanglement and correlation \cite{Friis13,Friis16,Sergey17}. In this context, the term `mode' shall not be confused with the modes emerging in the nuclear problem in quantum chemistry. The second route (discussed in Section \ref{sec:part}) is related more to first quantization and tries to define correlation and entanglement in the \emph{particle} picture.

\subsubsection{Mode/Orbital Picture}\label{sec:mode}
A natural tensor product structure emerges in the formalism of second quantization, facilitating a bipartition on the set of spin-orbitals. To explain this, let us fix a reference basis for the one-particle Hilbert space $\mathcal{H}_1$. We then introduce the corresponding fermionic creation ($f_i^{\dagger}$) and annihilation operators ($f_j$), fulfilling the fermionic commutation relations,
\begin{equation}
    \{ f^{(\dagger)}_i, f^{(\dagger)}_j \} = 0, \quad \{f^\dagger_i, f_j \} = \delta_{ij}. \label{anticommute}
\end{equation}
In the quantum information community the one-particle reference states are often referred to as \textit{modes}, or (lattice) sites by condensed matter physicists. Each spin-orbital or generally mode $i$ can be either empty or occupied by a fermion. In this picture, the quantum states are naturally represented in the occupation number basis. The respective \emph{configuration states}
\begin{equation}\label{config}
    |n_1,n_2,\ldots , n_d \rangle = (f^\dagger_1)^{n_1}(f^\dagger_2)^{n_2} \cdots (f^\dagger_d)^{n_d} |0\rangle
\end{equation}
with $n_1,n_2\ldots,n_d \in \{0,1\}$ form a basis for the Fock space $\mathcal{F}(\mathcal{H}_1)$.
Bipartitions naturally arise as separations of the set of modes $\{1,2,\ldots,d\}$ into two, let's say the first $m$ and the last $d-m$, leading to
\begin{eqnarray}\label{slitconfig}
\lefteqn{|n_1, \ldots, n_m, n_{m+1},\ldots , n_d \rangle} && \nonumber \\
        &\mapsto &|n_1, n_2, \ldots n_{m} \rangle_A \otimes |n_{m+1}, n_{m+2} , \ldots ,n_{d}\rangle_B.
\end{eqnarray}
The total Fock space $\mathcal{F}(\mathcal{H}_1)$ admits then the tensor product structure
\begin{equation}\label{splitFock}
    \mathcal{F}_{AB}\equiv \mathcal{F}(\mathcal{H}_1) = \mathcal{F}(\mathcal{H}_1^{(A)}) \otimes \mathcal{F}(\mathcal{H}_1^{(B)})\equiv \mathcal{F}_A \otimes \mathcal{F}_B,
\end{equation}
where $\mathcal{H}_1^{(A/B)}$ denotes the one-particle Hilbert space spanned by the first $m$ and last $d-m$ modes, respectively. Actually, any splitting of the one-particle Hilbert space into two complementary subspaces, $\mathcal{H}_1 = \mathcal{H}_1^{(A)}\oplus \mathcal{H}_1^{(B)}$, induces a respective splitting \eqref{splitFock} on the Fock space level. Moreover, such a decomposition of the total Fock space into two factors allows us to introduce mode reduced density operators $\rho_{A/B}$ for the respective mode subsystem $A/B$. They are obtained by taking the partial trace of the total state $\rho$ with respect to the complementary factor $\mathcal{F}_{B/A}$. Consequently, $\rho_{A/B}$ is defined as an operator on the local space $\mathcal{F}_{A/B}$ and in general does not have a definite particle number anymore.

It seems that we can now readily apply the common quantum information theoretical formalism referring to distinguishable subsystems (as it has been done in quantum chemistry during the past few years\cite{boguslawski2012entanglement,boguslawski2013orbital,boguslawski2015orbital}). Yet there is one crucial obstacle.
Not every Hermitian operator acting on a fermionic Fock space is a physical observable. For instance, nature does not allow one to coherently superpose even and odd fermion number states\cite{SSR}. The significance of this \emph{number parity superselection rule} (SSR) is rather obvious since its violation would  equivalently make superluminal signalling possible\cite{ZZ14,Friis16}, in contradiction to special relativity. The number parity SSR implies that the algebra of observables on any Fock space comprises only those operators which are block-diagonal with respect to the even and odd fermion number sectors, $\hat{A}= \hat{A}_{ee}+ \hat{A}_{oo}$\cite{Banuls07}.

While the number parity SSR has a deep physical origin and is well-known in other fields of physics since several decades\cite{SSR}, even additional, more restrictive SSRs could apply in practice. Those would not necessarily represent fundamental physical limitations but could also reflect the impossibility to experimentally realize specific Hermitian operators. In the context of our work, the number parity SSR needs to be replaced by the more restrictive \emph{particle number} SSR. Indeed, in quantum chemistry observables are particle number conserving since it is highly unlikely that fermion pairs emerge from vacuum fluctuations. Consequently, the local algebras $\mathcal{A}_{A/B}$ of observables comprise only those operators which are block-diagonal in all particle number sectors,
\begin{equation}\label{NSSR}
\hat{A}= \sum_{N\geq 0} P_N \hat{A} P_N,
\end{equation}
where $P_N$ denotes the projector onto the $N$-fermion sector $\mathcal{H}_N$.

As indicated below Definition \ref{def:sep}, it is still straightforward to define total correlation and entanglement in the presence of an SSR\cite{verstraete2003quantum,schuch2004nonlocal,vaccaro2008tradeoff}. One just needs to refer in Definition \ref{def:uncorr} to the correct physical algebras $\mathcal{A}_{A/B}$. In that case, more quantum states are uncorrelated since the algebras $\mathcal{A}_{A/B}$ are getting smaller (recall Definitions \ref{def:uncorr} and Eq.~\eqref{CorrMeasure}). Since the concept of entanglement (recall Definition \ref{def:sep} and Eq.~\eqref{EntMeasure}) refers to a notion of total correlation the same will be the case for it as well.
This also means that erroneously ignoring the number parity or particle number SSR would lead to an overestimation of the extractable total correlation and entanglement in concrete quantum systems.

\subsubsection{Particle Picture}\label{sec:part}
The formalism of first quantization seems to suggest another tensor product structure by exploiting the embedding
\begin{equation}
    \mathcal{H}_N \equiv  \wedge^N(\mathcal{H}_1) \leq \mathcal{H}_1^{\otimes N} ,
\end{equation}
of $N$-fermion Hilbert space into the one of $N$ distinguishable particles.
As already explained in the introduction, the problem is that the antisymmetry of $N$-fermion quantum states now erroneously would contribute to this particle correlation/entanglement. Or equivalently, not every Hermitian operator on $\mathcal{H}_1^{\otimes p}$ is a fermionic observable. %

Consequently, there is little doubt that a notion of correlation and entanglement between identical fermions cannot exist in the concise quantum information theoretical sense. Yet, there is well-defined alternative concept inspired by resource theory\cite{ResT19} which looks rather similar: One defines the configuration states as the distinguished resource-free states
\begin{defn}[Free States]\label{def:uncorrP}
A fermionic state $\rho $ is called free in the particle picture, if and only if it can be represented by a single configuration state, i.e., $\rho\equiv \ket{\Psi}\!\bra{\Psi} $ with,
\begin{equation}\label{config1st}
\ket{\Psi}= \ket{\phi_1} \wedge \ldots \wedge \ket{\phi_N}
\end{equation}
for some (orthonormal) one-fermion states/modes $\ket{\phi_1}, \ldots, \ket{\phi_N} \in \mathcal{H}_1$.
\end{defn}
\noindent Furthermore, in analogy to the separable states one defines
\begin{defn}[Quantum-Free States]\label{def:sepP}
A fermionic state $\rho$ is called quantum-free in the particle picture, if and only if it can be written as a mixture of free states.
\end{defn}
\noindent A few comments are in order. First, the definition of free and quantum-free states could be applied in the context of both fixed particle number ($N$-fermion Hilbert space) and flexible particle number (Fock space). Second, since the definitions of free and quantum-free states look rather similar to those of uncorrelated and non-entangled states, we denote the respective sets by $\mathcal{D}_0^{(p)}$ and $\mathcal{D}_{sep}^{(p)}$, respectively. The superindex `$p$' refers to the particle picture and similarly we will add in the following a superindex `$m$' to the corresponding sets in the mode/orbital picture (as introduced by Definitions \ref{def:uncorr}, \ref{def:sep}).

Measures for the nonfreeness and its quantum part can then be obtained by determining the minimal distances of a given quantum state $\rho$ to the sets $\mathcal{D}_0^{(p)}$ and $\mathcal{D}_{sep}^{(p)}$, respectively. Due to the close relation of this \emph{(quantum) nonfreeness} to the concepts of total (quantum) correlation we denote the respective measure by $(E^{(p)})\, C^{(p)}$.
Actually, the nonfreeness was first introduced by Gottlieb and Mauser\cite{gottlieb2005new,gottlieb2007properties,Gottlieb14,Gottlieb15} and they  observed\cite{gottlieb2007properties} that using the quantum relative entropy as distance function leads to an analytic formula  (referring to a Fock space-related Definition \ref{def:uncorrP}),
\begin{equation}\label{PartCorr}
    C^{(p)}(\rho) = S(\rho_1)+ S(\mathbbm{1}-\rho_1)-S(\rho).
\end{equation}
In this formula, the 1RDM $\rho_1$ of $\rho$ is trace-normalized to the particle number $N$. In case of pure total states $\rho$, $S(\rho)$ vanishes and the nonfreeness $C^{(p)}$ is nothing else than the particle-hole symmetrized von Neumann entropy of the 1RDM. Since this nonfreeness has a beautiful geometric meaning, the chances for discovering an underlying operational meaning might be better than for the quantity $S(\rho_1)$ as used in most works so far (see, e.g., Refs.~\onlinecite{Ziesche97b,huang2005entanglement,Schoutens08}).

Deriving an explicit analytic expression for the quantum part $E^{(p)}(\rho)$ of the nonfreeness seems to be a rather hopeless task again (as for the entanglement of mixed states in general). It is thus quite remarkable that at least for the case of two fermions in a four-dimensional one-particle Hilbert space an analytic procedure has been found\cite{schliemann2001quantum} (which, however, does not involve the quantum relative entropy and instead is based on a so-called convex-roof construction):
In a first step, one determines the spectral decomposition of the given two-fermion density operator $\rho$ on $\wedge^2[\mathcal{H}_1]$ (here the respective eigenvalues are absorbed into the states $\ket{\Psi_i}$),
\begin{equation}
\rho= \sum_{i=1}^6 \ket{\Psi_i}\!\bra{\Psi_i}.
\end{equation}
By introducing an arbitrary reference basis for $\mathcal{H}_1$, one determines for all six contributions $\ket{\Psi_i}$ the antisymmetric expansion
matrices $w^{(i)}$,
\begin{equation}
    |\Psi_i\rangle = \sum_{a,b=1}^4 w^{(i)}_{ab} f^\dagger_a f^\dagger_b |0\rangle.
\end{equation}
Those are then used to calculate for $i,j=1,\ldots,6$
\begin{equation}
    K_{ij} = \sum_{a,b,c,d=1}^4 \epsilon^{abcd} w^{(i)}_{ab} w^{(j)}_{cd} \label{C_SL}.
\end{equation}
The quantum nonfreeness eventually follows as\cite{schliemann2001quantum}
\begin{equation} \label{PartEnt}
    E^{(p)}(\rho) = 2 \max_i |\kappa_i| - \Tr[|K|],
\end{equation}
where $\{\kappa_i\}$ are the eigenvalues of the matrix $K\equiv(K_{ij})$ and $\Tr[|K|]= \sum_{i=1}^{6}|\kappa_i|$.

Equipped with the measures $C^{(p)}(\rho)$ and $E^{(p)}(\rho)$ we can quantify in the following how close a given state $\rho$ is to the family $\mathcal{D}^{(p)}_0$ of configuration states and to its convex hull $\mathcal{D}_{sep}^{(p)}$ given by their classical mixtures.

\section{Effective Model and Qualitative Resolution of the Correlation Paradox}\label{sec:diss}
In this section we first recall the correlation paradox and then resolve it for the hydrogen molecule in the dissociation limit in a \emph{qualitative} way for infinite separation $r$. Moreover, to consider finite (large) distances $r$ between both nuclei we introduce and solve the Hubbard dimer model as an effective model for the dissociation limit.

\subsection{Correlation paradox}\label{sec:paradox}
Let us first recall the correlation paradox from a more general point of view. Whenever identical fermions do not interact, solving the $N$-particle Schr\"odinger equation simplifies to an effective one-fermion problem. Indeed, for any Hamiltonian $\hat{H}\equiv \sum_{i,j=1}^d h_{ij} f_i^\dagger f_j $ one just needs to diagonalize the Hermitian matrix $(h_{ij})$, leading to $\hat{H} \equiv \sum_{\alpha=1}^d h_\alpha \hat{n}_\alpha$ with some one-particle solutions $\ket{\alpha}$. The respective $N$-fermion eigenstates follow as configuration states $\ket{\alpha_1,\ldots,\alpha_N}\equiv \ket{\alpha_1}\wedge \ldots \wedge\ket{\alpha_N}$ obtained by distributing the $N$ fermions into $N$ different spin-orbitals $\ket{\alpha_i}$. Having said this, how can a non-degenerate fermionic ground state not take the form of a single configuration state in a limit process which marginalizes the interaction between the fermions? The existence of exactly such processes can be seen as paradoxical in that sense.

The dissociation limit of molecules often gives rise to such paradoxical situations which play an important role in the context of the general electron correlation problem\cite{Low58,Pop76,Bart78,Pop87,Mazz12}.
They are of course well-understood in quantum chemistry, in particular on a qualitative level. For instance, it is rather obvious that those paradoxical situations require the closing of the excitation gap $\Delta E(r)$ and at the limit $r\rightarrow \infty$ the system needs to have several configuration states as degenerate ground states. For very large but not infinite separation distances $r$ between the nuclei, those configurations are then typically superposed to form the non-degenerate correlated ground state. From the most rudimentary point of view, the paradox could therefore be resolved by just referring to the excitation gap $\Delta E(r)$ which reduces to zero at least as fast as the electron-electron interaction energy vanishes.

Yet, there is more to be said. For instance, why would one like to construct a measure for correlation\cite{Ziesche97} which vanishes for the dissociated hydrogen ground state \eqref{PsiH2inf} despite the fact that the latter cannot be written as a single configuration state? It seems to us that there are partly self-contradicting definitions in place for what ``correlation'' actually might or should be:
On the one hand, a state is considered as being ``uncorrelated'' if it takes the form of as a single configuration state. On the other hand, one observes that the electron-electron interaction vanishes in the dissociation limit despite the fact that the ground state is not a configuration state. This apparent contradiction is based on a confusion between the notion of total correlation and the concept of correlation functions. Furthermore, how does the dissociated ground state \eqref{PsiH2inf} compare to the uncorrelated degenerate configuration states emerging at the limit $r\rightarrow \infty$ in terms of its robustness to perturbations? We will provide an answer to the latter question in Section \ref{sec:resol}. To be more specific, we illustrate and prove that thermal noise due to finite, possibly even just infinitesimally low, temperature $T$ will destroy the quantum correlations beyond a critical separation distance $r_{\mathrm{crit}}$($T$) entirely.
This rationalizes that ``correlation'' vanishes in the dissociated ground state in the sense that this perception is correct provided the presence of some (possibly infinitesimally low) temperature $T>0$. These considerations which are made precise in Section \ref{sec:resol}
reveal a conceptually new characterization of static and dynamic correlation in ground states by relating them
to the (non)robustness of correlation with respect to thermal noise.

\subsection{Hubbard dimer as an effective model}\label{sec:hubbard}
From a general point of view, the realization of the dissociation limit of the hydrogen molecule (or any other molecular system) in the laboratory requires the coupling of the molecule to another system. To present our theoretical argument on the (in)stability of correlation/entanglement with respect to thermal noise in the cleanest fashion we consider an experimental procedure which accesses the nuclei directly to moves them apart. In that sense, it also freezes the nuclear (vibrational) degrees of freedom and the Born-Oppenheimer approximation with a separation distance $r$ of both nuclei will be assumed. To discuss such realizations of the dissociation limit of the hydrogen molecule we thus begin with the electronic Hamiltonian, i.e., we consider two interacting electrons in the Coulomb potential generated by two nuclei separated by a distance $r$. Choosing large basis sets of atomic orbitals centered around both nuclei would allow one to obtain highly accurate descriptions of the behavior and the properties of the hydrogen molecule. Yet, in our case we restrict ourselves to very low temperatures and thus only the $1s$ orbital needs to be taken into account for capturing the main effects.
This is due to the fact that the energy difference between the 1s and the higher shells significantly exceeds the energy scale of the electron-electron energy in atoms. After all, this approximation is getting exact in the limit of arbitrarily large separation distances $r$
since then the two electrons are getting arbitrarily far separated (and in particular the probability of finding them at the same nucleus tends to zero).

As a consequence, we can study the most relevant aspects of the dissociation limit of the hydrogen molecule in the Hubbard dimer model\cite{alvarez2001hubbard,chiappe2007hubbard}. This (and after all our initial choice of a simple two-electron system) will allow us to illustrate all relevant quantum information theoretical aspects without getting deflected by highly involved descriptions of correlated ground states.
From a general point of view, the Hubbard dimer is one of the simplest models for interacting fermions, while already exhibiting rich physical properties. It consists of two lattice sites ($L$ and $R$) corresponding to the 1s orbitals centered at both nuclei and the underlying Hamiltonian takes the form
\begin{equation}\label{Hubbard}
    \hat{H} = - t\!\sum_{\sigma = \uparrow, \downarrow} (f^\dagger_{L\sigma} f_{R\sigma } + f^\dagger_{R\sigma} f_{L\sigma} ) + U \sum_{i=L,R}\hat{n}_{i\uparrow}\hat{n}_{i\downarrow}.
\end{equation}
Here, $t\geq 0$ describes the hopping between the both nuclei/sites, $U>0$ represents the on-site repulsion (originating from the Coulomb interaction between two electrons in a 1s shell) and $\hat{n}_{L/R}$ denotes the particle number operator at the left/right site. Since the eigenstates of \eqref{Hubbard} depend only on the ratio $t/U$  we set in the following $U\equiv 1$. Moreover, the hopping $t$ decays exponentially as function of $r$, in agreement with the overlap of two 1s atomic orbitals separated by a distance $r$\cite{krauss1965interaction}. Depending on the context, we will choose in the following either $r$ or $t=e^{-r}$ as the parameter of the system.

\subsection{Qualitative Resolution of the Correlation Paradox}\label{sec:Resolqual}
The fully dissociated ground state of the Hubbard dimer model and the hydrogen molecule, respectively, follows as the singlet state \eqref{PsiH2inf}. Independent of which reference orbitals are chosen that state can never be written as a configuration state and is also not close to any. In general, this can easily be confirmed in terms of the decreasingly ordered natural occupation numbers: The more they deviate from the spectrum $(1,\ldots,1,0,\ldots,0)$ of a configuration state, the larger the distance of the quantum state to the closest configuration state (see also Lemma 3.6.1 in Ref.~\onlinecite{CSthesis} for more details). In our case they follow as $1/2(1,1,1,1)$.
Since at the same time both electrons are infinitely far separated the dissociation limit of the hydrogen molecule represents an example for a correlation paradox. According to Section \ref{sec:paradox}, this requires in particular that at the limit $r\rightarrow \infty$, or equivalently $t=0$ in the Hubbard dimer, the system has degenerate configuration states as ground states. That is the case indeed and those configuration states follows as $f_{L\sigma}^\dagger f_{R\sigma'}^\dagger \ket{0}\equiv \ket{L\sigma,R\sigma'}$, $\sigma,\sigma' =\uparrow,\downarrow$. If we consider the same system at finite temperature $T>0$, however, the situation changes considerably since the system is not in its ground state anymore. Instead, as long as the thermal energy $k_B T$ exceeds significantly the excitation gap $\Delta E(r)$, $ k_B T\gg \Delta E(r)$ (which can be achieved for each $T>0$ by choosing $r$ sufficiently large),
there will be an equal weighted mixing of the singlet ground state with the three triplet excited states.
The respective quantum state of the system thus follows as
\begin{equation}
\rho(T,r) \approx \frac{1}{4} \sum_{\sigma,\sigma' =\uparrow,\downarrow}\! \ket{L\sigma,R\sigma'}\!\bra{L\sigma,R\sigma'}.
\end{equation}
This state is still not a configuration state but a classical mixture of them. This means that from a particle picture's viewpoint thermal noise destroys quantum nonfreeness and turns it into classical nonfreeness. To also quantify the mode correlation and entanglement, we observe that
$\rho(T,r)$ factorizes in the mode/orbital picture (i.e., second quantization) according to $\rho(T,r)= \frac{1}{4}(\sum_{\sigma=\uparrow,\downarrow} \ket{L\sigma}\!\bra{L\sigma}) \otimes (\sum_{\sigma'=\uparrow,\downarrow} \ket{R\sigma'}\!\bra{R\sigma'})$. Hence, even if we were neglecting the particle number superselection rule, the dissociated thermal state is not mode entangled and even not mode correlated with respect to the left/right split of the system.

\subsection{Diagonalization of the Hubbard Dimer}\label{sec:HubbardSol}
To diagonalize the Hamiltonian \eqref{Hubbard} of the Hubbard dimer it is instructive to exploit
its spin symmetries and the reflection symmetry $L \leftrightarrow R$. Those manifest themselves in the form of the total spin $S$, the magnetization $M$ along the $z$-axis and the refection parity $p$ as good quantum numbers. The corresponding eigenvalue problem decouples according to
\begin{equation}
\hat{H} = \bigoplus_{S=0}^1 \bigoplus_{M=-S}^S \bigoplus_{p=\pm} \hat{H}_{S,M,p}.
\end{equation}
As a matter of fact, this almost completes the diagonalization and it remains to diagonalize $\hat{H}_{0,0,-}$ on the corresponding two-dimensional space $\mathcal{H}_{0,0,-}$. The details of those calculations are presented in Appendix \ref{spectrum} and we just present here the well-known results for the six eigenenergies \cite{hasegawa2011thermal} (with $U\equiv 1$)
\begin{eqnarray}\label{dimerspectrum}
&&E_0 = \frac{1}{2} - \sqrt{\frac{1}{4}+4t^2}\,, \quad E_1 = E_2 = E_3 = 0, \nonumber \\
&&E_4 = 1, \quad E_5 = \frac{1}{2} + \sqrt{\frac{1}{4}+4t^2}.
\end{eqnarray}
It is crucial to notice that the ground state is always non-degenerate and the first excited energy corresponds to the threefold degenerate  triplet states. The energy spectrum \eqref{dimerspectrum} is also shown in Figure \ref{HubbardSpectrum} (recall $t = \exp(-r)$) and the corresponding six eigenstates are listed in Appendix \ref{spectrum}.
\begin{figure}[h!]
    \centering
    \includegraphics[scale=0.22]{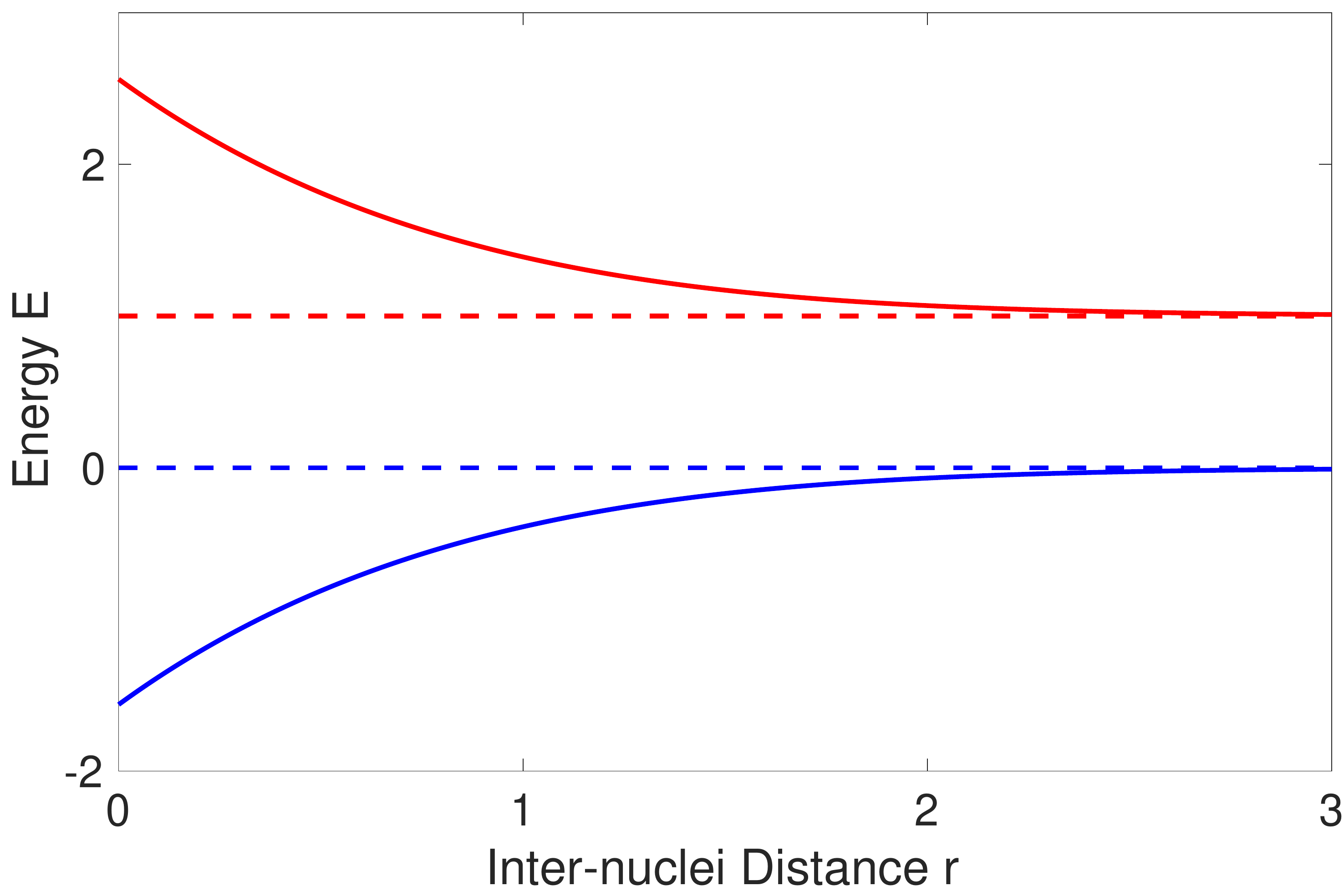}
    \caption{Energy spectrum \eqref{dimerspectrum} of the Hubbard dimer \eqref{Hubbard} in dimensionless units ($U$ and the Bohr radius are set to one). All energy levels are non-degenerate except for the first excited level (blue dashed), where three triplet states reside.}
    \label{HubbardSpectrum}
\end{figure}
In particular this confirms the closing of the excitation gap for $r\rightarrow \infty$, as described by $\Delta E(r) \sim 4 t^2=4 e^{-2 r}$.

At temperature \(T=0\) the system takes the energetically favorable ground state,
\begin{eqnarray}\label{ground}
    |\Psi_0(r)\rangle &=& \frac{a(r)}{\sqrt{2}}\big(f^\dagger_{L\uparrow}f^\dagger_{R\downarrow} -  f^\dagger_{L\downarrow} f^\dagger_{R\uparrow}\big)\ket{0} \nonumber \\
    && + \frac{b(r)}{\sqrt{2}} \big(f^\dagger_{L\uparrow}f^\dagger_{L\downarrow} -  f^\dagger_{R\downarrow} f^\dagger_{R\uparrow}\big) \ket{0},
\end{eqnarray}
where the coefficients $a(r)$ and $b(r)$ are functions of the inter-nuclear distance $r$ (explicit expressions can be found in Appendix \ref{spectrum}). In particular, one has $a(r)=\sqrt{1-b^2(r)}$ and $b(r)\sim 2 \,t = 2 e^{-r}$ for $r \rightarrow \infty$. The latter confirms
that the probability of finding both electrons at the same site/nucleus tends to zero for large separation distances $r$ and small hopping rates $t$, respectively.
Consequently, at the limit \(r \rightarrow \infty\), the ground state follows indeed as
\begin{equation}
    |\Psi_0(r=\infty)\rangle = \frac{1}{\sqrt{2}}(f^\dagger_{L\uparrow}f^\dagger_{R\downarrow} - f^\dagger_{L\downarrow} f^\dagger_{R\uparrow})|0\rangle,
\end{equation}
which is \textit{not} a configuration state.

At finite temperature, the state of interest is the thermal Gibbs state (we set for simplicity $k_B \equiv 1$),
\begin{equation}
    \rho(T,r) = \frac{1}{Z(T,r)} e^{-\hat{H}(r)/ T} , \label{Gibbs}
\end{equation}
where $Z(T,r)\equiv \Tr\left[e^{-\hat{H}(r)/T}\right]$ is the partition function.
In addition to the standard Boltzmann-Gibbs statistics which we use here, there have also been proposals of other distributions for systems of non-extensive size\cite{salinas1999special,hasegawa2011thermal}. Although it is somewhat debatable to say which statistics is more appropriate for a small system like ours, we shall stick to the Boltzmann-Gibbs distribution and use the thermal equilibrium state as defined in Eq.~\eqref{Gibbs}.

\section{Correlation and Entanglement Analysis}\label{sec:resol}
In this section we determine the mode entanglement $E^{(m)}$, mode correlation $C^{(m)}$, nonfreeness $C^{(p)}$ and its quantum part $E^{(p)}$ of the thermal equilibrium state \eqref{Gibbs} of the Hubbard dimer. To resolve the correlation paradox of the dissociation limit by referring to thermal noise, we consider in particular the regime of low temperatures $T$ and large separation distances $r$.

\subsection{Mode/Orbital Picture}\label{sec:resolmode}
We first consider the total mode correlation between the left and right side/nucleus. As explained in Section \ref{sec:mode} this
means to split the one-particle Hilbert space according to $\mathcal{H}_1 = \mathcal{H}_1^{(L)} \oplus \mathcal{H}_1^{(R)}$ where $\mathcal{H}_1^{(i)}$ is spanned by $\{\ket{i\!\uparrow},\ket{i\!\downarrow}\}$, $i=L/R$. This implies the decomposition $\mathcal{F}= \mathcal{F}_L \otimes \mathcal{F}_R$ into left and right mode subsystems, $\mathcal{F}_{L/R}\equiv \mathcal{F}[\mathcal{H}_1^{(L/R)}]$ . If no particle number SSR applied, the mode correlation would be given by the quantum mutual information \eqref{CorrMeasure}.  Yet, in our case the local observables fulfill the particle number SSR \eqref{NSSR} on both sides. The comparison of the total state $\rho$ with the local ones, $\rho_L \otimes \rho_R$, is relative to the algebra of local observables. Consequently, those blocks of $\rho$ which cannot affect any local measurement must be cut out\cite{Wiseman03}, leading to
\begin{equation}\label{EntMeasureNSSR}
\tilde{\rho} \equiv \sum_{N',N''=0}^2 P_{N',N''} \,\rho\, P_{N',N''}.
\end{equation}
Here $P_{N',N''}\equiv P_{N'} \otimes P_{N''}$ denotes the projector onto the sectors with $N'$ particles on the left and $N''$ particles on the right side. The mode correlation then follows as the quantum mutual information of the particle number SSR-adapted state \eqref{EntMeasureNSSR}.

The calculation of the mode entanglement is typically computationally demanding. Even in our case with a decomposition of the total space into two just four-dimensional subspaces, $\mathcal{F}= \mathcal{F}_L \otimes \mathcal{F}_R$, the minimization with respect to $\sigma$ in \eqref{EntMeasure} involves initially 256 parameters. Yet, our total state \eqref{Gibbs} exhibits many symmetries which can be taken into account to reduce the number of parameters. Let us briefly outline how this works (for more details of those technical concepts we refer the reader to Refs.~\onlinecite{vollbrecht2001entanglement,bartlett2003entanglement} and our forthcoming article \cite{LexinQIT}).
The underlying key insight is, provided some assumptions are met, that the closest separable state $\sigma$ to a state $\rho$ would have the same local unitary symmetries enjoyed by $\rho$. To explain this, suppose $\sigma$ is the closest separable state to $\rho$, and $\rho$ is invariant under a local unitary transformation $U$ belonging to a group $G$, then
\begin{equation}
    \begin{split}
        S(\rho||\sigma) = S(U\rho U^\dagger || \sigma ) = S(\rho || U^\dagger \sigma U),
        \quad \forall U \in G.
    \end{split}
\end{equation}
Consequently, $U^\dagger \sigma U$ is also a minimizing state. By referring to the convexity of the relative entropy we obtain
\begin{equation}
    \begin{split}
        S(\rho||\sigma) \geq S(\rho || \mathcal{T}_G(\sigma)). \label{twirl}
    \end{split}
\end{equation}
The \emph{twirl} $\mathcal{T}_G$ with respect to the group $G$ is defined as
\begin{equation}
    \mathcal{T}_G (\sigma) = \begin{cases}
    \frac{1}{|G|} \sum_{U \in G} U^\dagger \sigma U, \quad &G \mathrm{\:\, \mbox{discrete}},
    \\
    \int_G\,  U^\dagger \sigma U \, \mathrm{d}\mu(U), \quad &G \mathrm{\:\, \mbox{non-discrete}}.
    \end{cases}
\end{equation}
Here in the second case (e.g., with $G$ a Lie group), the integration is performed with respect to the so-called Haar measure.
The result of twirling $\sigma$ is that $\mathcal{T}_G(\sigma)$ is by construction $G$-symmetric, namely it commutes with all elements of $G$. It then follows that the separable state $\sigma$ closest to $\rho$ is $G$-symmetric as well.

The Gibbs state $\rho$ defined in Eq.~\eqref{Gibbs} enjoys many local symmetries (recall Section \ref{sec:hubbard}).
Exploiting all of them leads to a drastic reduction of the relevant parameters involved in \eqref{EntMeasure}. The respective
minimization process can then easily be performed on a computer.
A mathematically rigorous and detailed derivation of those concepts in an even more general setting can be found in our forthcoming work\cite{LexinQIT}.

\begin{figure}[h!]
    \centering
    \includegraphics[scale=0.26]{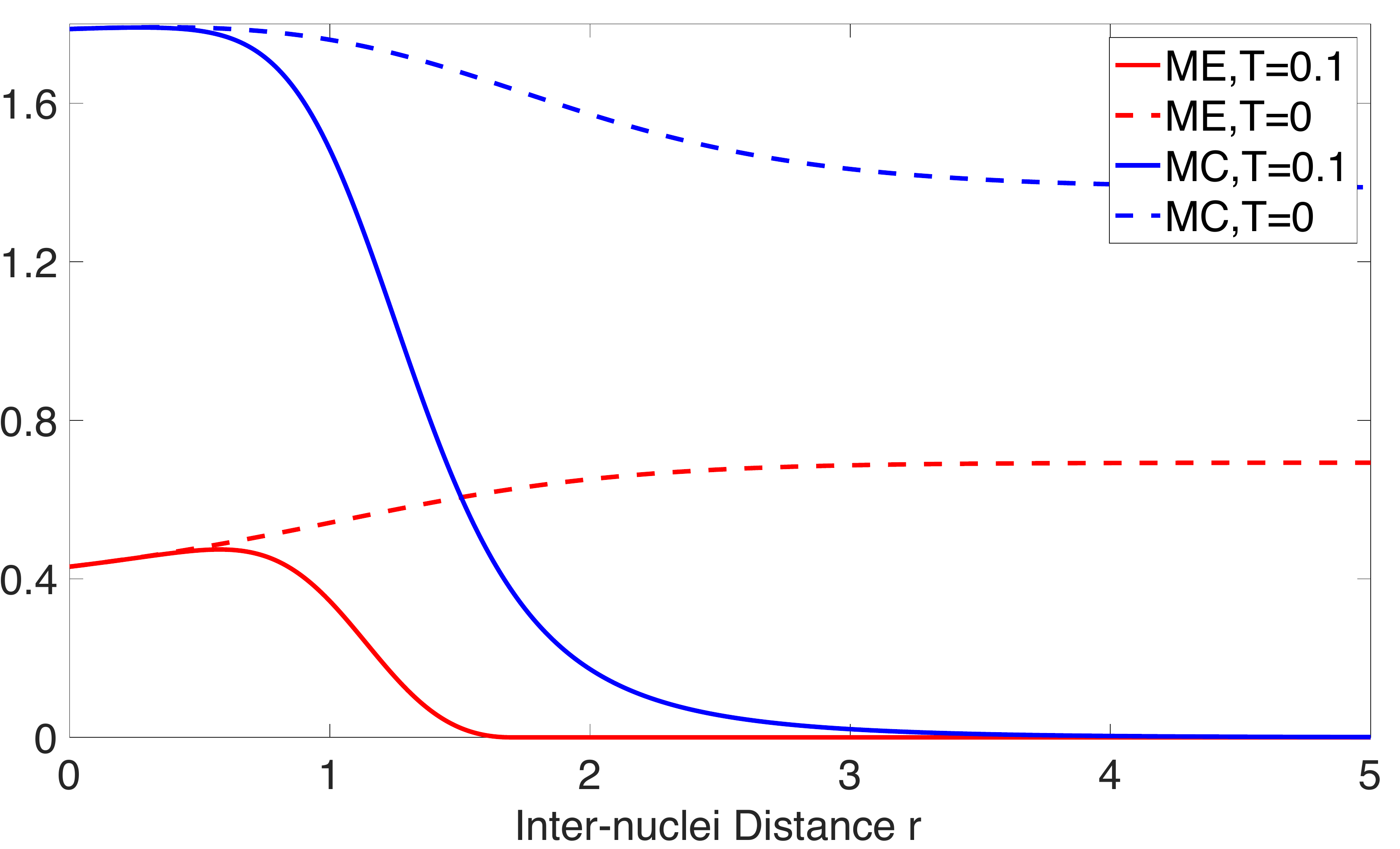}
    \caption{Mode correlation (MC, blue) and mode entanglement (ME, red) plotted for the Gibbs state (solid) at finite temperature $T=0.1$ and the ground state (dashed), the equilibrium state at $T=0$, with particle number superselection rule.}
    \label{fig:ModeCorr}
\end{figure}

We are now in a position to calculate the mode correlation and mode entanglement in the Gibbs state in Eq.~\eqref{Gibbs} for all temperatures $T$ and all separation distances $r$. The respective results for the cases $T=0, 0.1$ are presented
in Figure \ref{fig:ModeCorr}.
First, we observe that both mode correlation and mode entanglement in the Gibbs state with $T=0.1$ and the ground state ($T=0$) coincide at smaller distances $r$. This is due to the fact that for small $r$ the energy gap $\Delta E(r)$ between the ground state and the first excited states is much larger than the thermal energy scale $k_B T$ such that both states essentially coincide (the contribution of the excited states to the Gibbs ensemble are exponentially suppressed according to Eq.~\eqref{Gibbs}).
Second, the presence of a correlation paradox is confirmed since the mode correlation (blue dashed) and mode entanglement (red dashed) of the ground state remain finite even at the dissociation limit. Third, for finite temperature, this is quite different. When the inter-site distance $r$ becomes larger, and the gap $\Delta E(r)$ between ground state and first excited state closes, both correlation (blue solid) and entanglement (red solid) at finite temperature start to deviate more from the ground state ones. They get smaller and smaller, and they are eventually completely wiped out at the dissociation limit. This asymptotic behavior at $r \rightarrow \infty$ is present at any finite temperature $T>0$, regardless of how small it is. In particular, this means that the mode correlation of the ground state is highly unstable against thermal noise, and finite mode entanglement or mode correlation at the dissociation limit can \textit{never} be observed in a laboratory.

Remarkably, in Figure \ref{fig:ModeCorr} the mode entanglement in the Gibbs state drops to zero already at a \emph{finite} distance, $r_{crit}^{(m)}(T=0.1)=1.70$, unlike the usual asymptotic behavior of correlation. In other words, for any temperature $T$, there exists a minimal distance $r_{\mathrm{crit}}^{(m)}(T)$ beyond which the mode entanglement vanishes entirely. Such a decaying  behavior of the entanglement, sometimes referred to as ``sudden death'', is not uncommon in quantum systems\cite{yu2009sudden}, and is a unique feature of quantum correlation. Fascinating as it is, this finite parameter point at which the entanglement vanishes is nothing mysterious if one considers the geometric picture as shown in Figure \ref{fig:states}: The Gibbs state $\rho(T,r)$ simply entered the convex set of separable states as the inter-nuclei distance $r$ increases. In fact, because of this, the point $r_{\mathrm{crit}}^{(m)}(T)$ is of course independent of the measure employed for quantifying the entanglement.
\begin{figure}[h!]
    \centering
    \includegraphics[scale=0.26]{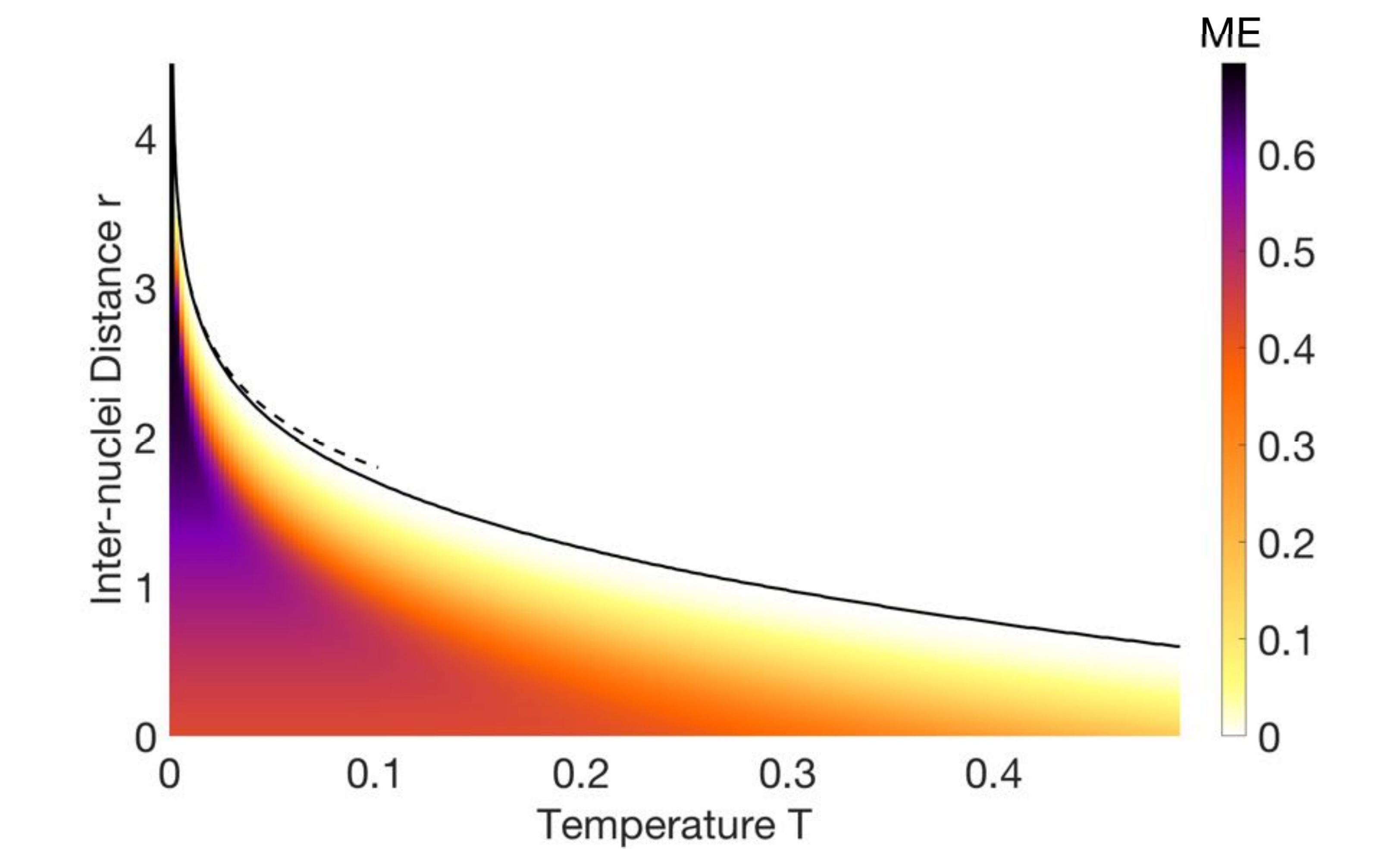}
    \caption{Mode entanglement (ME) as a function of temperature $T$ and inter-site distance $r$. It vanishes entirely above the black curve  $r_{\mathrm{crit}}^{(m)}$. The dashed line represents the asymptotic result \eqref{rm} for small $T$.}
    \label{fig:ME_sudden}
\end{figure}

To see how temperature affects this phenomenon, we present the mode entanglement in Figure \ref{fig:ME_sudden} as a function of the temperature $T$ and the inter-nuclei distance $r$. The critical distance $r_{\mathrm{crit}}^{(m)}(T)$ is shown as black curve.
For all parameter points $(T,r)$ above the black curve the mode entanglement vanishes, while it is finite for all points below it. As the temperature increases, the minimum distance $r_{\mathrm{crit}}^{(m)}(T)$ required to disentangle the left and right side becomes smaller. When $T \rightarrow 0$, the Gibbs state $\rho(T,r)$ approaches the ground state $\ket{\Psi_0}\!\bra{\Psi_0}$, and $r_{\mathrm{crit}}^{(m)}$ approaches infinity. In fact, the divergence of $r_{\mathrm{crit}}^{(m)}$ at small $T$ is logarithmic,
\begin{equation}
    r_{\mathrm{crit}}^{(m)}(T) = - \frac{1}{2} \log(T) + c_0 +c_1 T+\mathcal{O}(T^2), \quad T \rightarrow 0, \label{rm}
\end{equation}
where $c_0\equiv \log(2) - \frac{1}{2} \log(\log(3))$, $c_1\equiv -\frac{1}{2}(1+\log (3))$ are constants.
This asymptotic result is shown as dashed line in Figure \ref{fig:ME_sudden} and its derivation is included in Appendix \ref{divergence} for the interested readers.

\subsection{Particle Picture}\label{sec:resolpart}
To determine the nonfreeness we just need to calculate the 1RDM of the state \eqref{Gibbs} and plug it into the formula \eqref{PartCorr}. To calculate the quantum nonfreeness we can resort to the analytic procedure outlined in Section \ref{sec:part} since our model consists indeed of two fermions and a four-dimensional one-particle Hilbert space. We also would like to recall that in contrast to the other measures employed in our work the respective measure \eqref{PartEnt} for the quantum nonfreeness is not of the form \eqref{EntMeasure}. It namely does not involve the relative entropy as a distance function and is based on a so-called convex roof construction instead. Nonetheless, the used measure for the quantum nonfreeness quantifies how close a state is to the convex set $\mathcal{D}^{(p)}_{sep}$ given as the convex hull of single configuration states.

\begin{figure}[h!]
    \centering
    \includegraphics[scale=0.26]{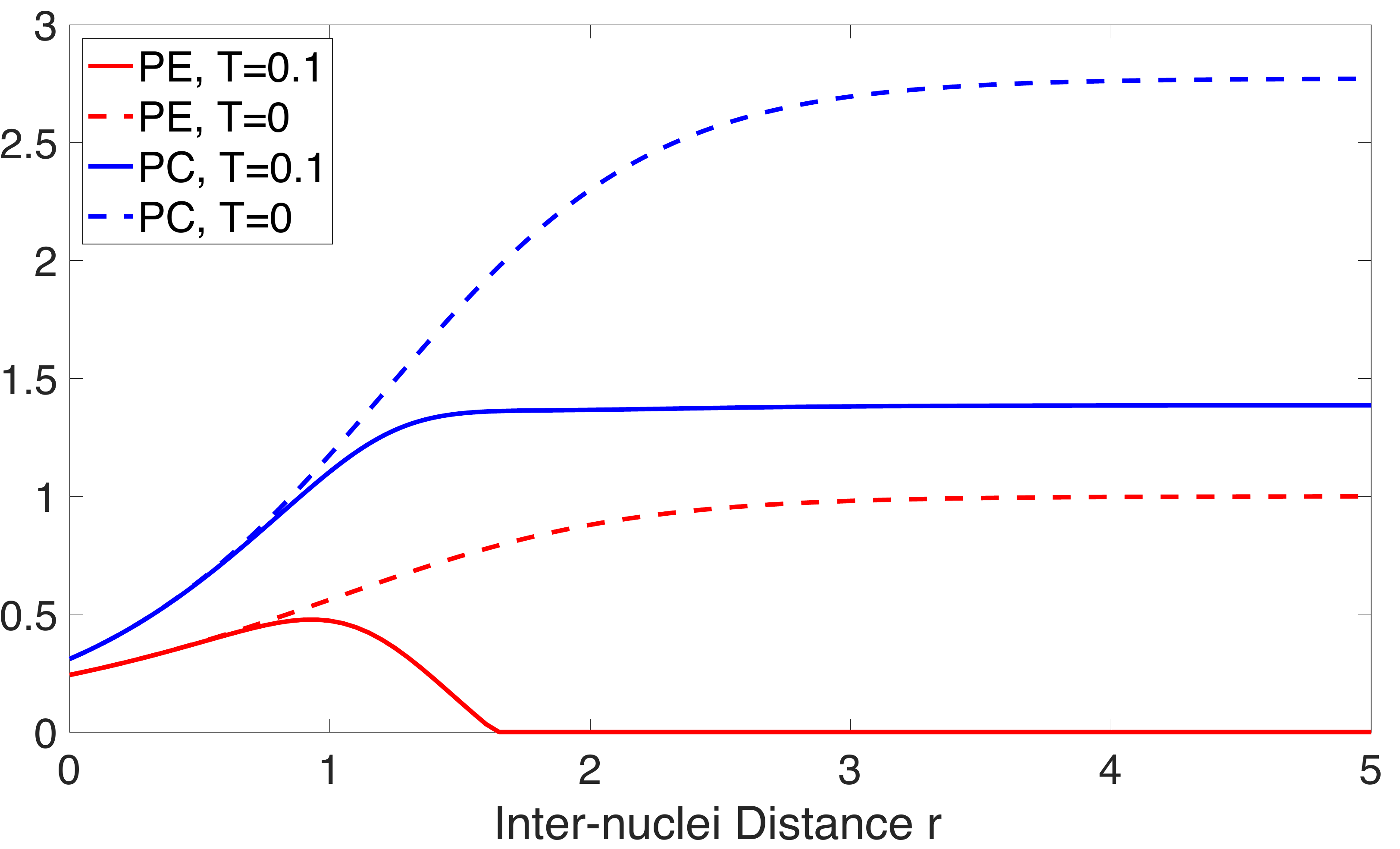}
    \caption{Nonfreeness (NF, blue) and quantum nonfreeness (QNF, red) as a function of inter-site distance. The correlation and entanglement in the ground state at zero temperature are also plotted (in dashed line).}
    \label{fig:ParticleCorr}
\end{figure}

In Figure \ref{fig:ParticleCorr} we present the nonfreeness \eqref{PartCorr} (blue) and its quantum part \eqref{PartEnt} (red) for the Hubbard dimer. 
The results for finite temperature $T=0.1$ are represented by the solid lines, and the dashed lines are reserved for the ground state Eq.~\eqref{ground} ($T=0$). As already discussed in the previous section, the Gibbs state at sufficiently low temperature approximately coincides with the ground state for smaller $r$, and therefore the (quantum) nonfreeness of both states approximately coincide as well. Things become very interesting as the two nuclei move further apart. First of all, the nonfreeness is reduced by introducing a small temperature, but remains finite at the dissociation limit. To be more specific, we already know that for any finite $T>0$, the Gibbs state approximates in the limit $r\rightarrow \infty$ better and better an equally weighted classical mixture of four configuration states,
\begin{equation}\label{PsiMix}
\rho(T,r) \approx \frac{1}{4}\sum_{\sigma,\sigma'=\uparrow/\downarrow} \ket{L\sigma,R\sigma'}\!\bra{L\sigma, R\sigma'}.
\end{equation}
This is also reflected by the fact that the 1RDM is perfectly mixed, $\rho_1=\frac{1}{2}\mathbbm{1}_4= \frac{1}{2}\sum_{i=L/R}\sum_{\sigma=\uparrow/\downarrow} \ket{i\sigma}\!\bra{i\sigma}$. This means that it is equally probable to find an electron on left or right, which has spin up or down. Moreover, as it can directly been inferred from the purely classical mixture \eqref{PsiMix} of configuration states, the quantum part of the nonfreeness decays to zero as we increase the inter-nuclei distance $r$. Remarkably, also the quantum nonfreeness in the Gibbs state experiences a ``sudden death'' as the mode entanglement, at a critical distance $r_{\mathrm{crit}}^{(p)}(T=0.1)=1.65$. As pointed out before, this phenomenon is a unique feature of quantum correlation, and it emphasizes that the quantum nonfreeness \eqref{PartEnt} captures something truly non-classical.
\begin{figure}[h!]
    \centering
    \includegraphics[scale=0.26]{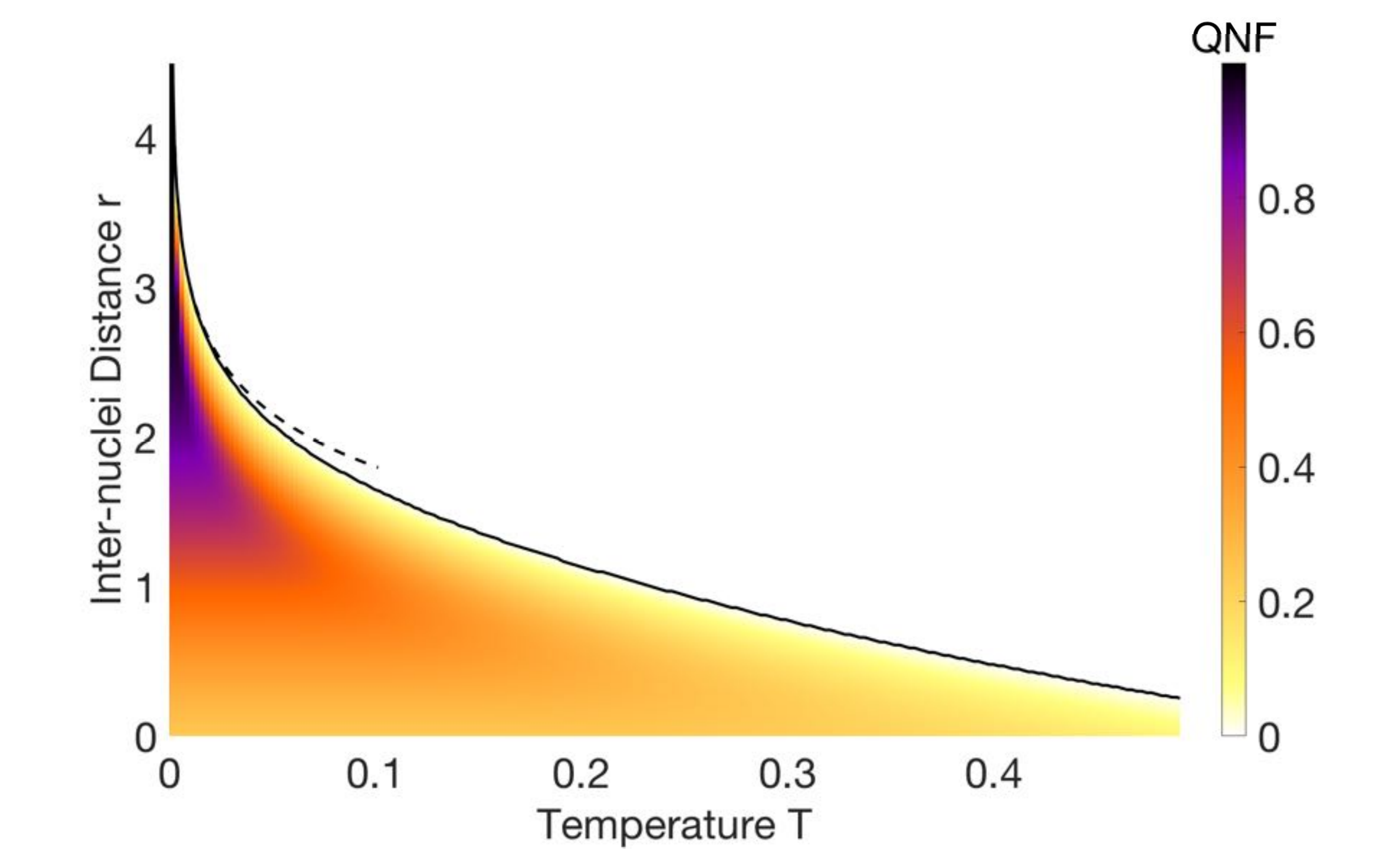}
    \caption{Quantum nonfreeness (QNF) as a function of temperature $T$ and inter-site distance $r$. It vanishes entirely above the black curve  $r_{\mathrm{crit}}^{(p)}$. The dashed line represents the asymptotic result \eqref{PartEntDiv} for small $T$.}
    \label{fig:SuddenDeath}
\end{figure}

To see how temperature affects the destruction of the quantum nonfreeness, we present the latter as a function of both distance and temperature in Figure \ref{fig:SuddenDeath}. The black line depicts $r_{\mathrm{crit}}^{(p)}$ as a function of $T$.
As the temperature increases, the minimum distance needed to kill the entanglement entirely is lowered. Similarly, the critical separation $r_{\mathrm{crit}}^{(p)}$ diverges logarithmically at small temperature,
\begin{equation}
    r_{\mathrm{crit}}^{(p)}(T) = - \frac{1}{2} \log(T) + d_0 +d_1 T+\mathcal{O}(T^2), \quad T \rightarrow 0, \label{PartEntDiv}
\end{equation}
where $d_0\equiv \log(2) - \frac{1}{2} \log(\log(3))$, $d_1\equiv -\frac{1}{2}(2+\log (3))$ are constants.
This asymptotic result is shown as dashed line in Figure \ref{fig:SuddenDeath} and its derivation is included in Appendix \ref{divergence} for the interested readers.

In the form of those results referring to the particle picture we have resolved the correlation paradox in the dissociation limit in the most concise way: For any finite temperature $T$, regardless of how close to zero it might be, there always exists a finite separation distance $ r_{\mathrm{crit}}^{(p)}(T)$ beyond which the quantum state $\rho(T,r)$ of the system does not contain quantum nonfreeness anymore. Instead, $\rho(T,r)$ is given as a purely classical mixture of configuration states. In particular, this means that the quantum nonfreeness in the ground state of the hydrogen molecule is highly unstable against thermal noise, and finite quantum nonfreeness at the dissociation limit can \textit{never} be observed in a laboratory.

\section{Correlation Paradox of the Generalized Dissociation Limit}\label{sec:general}
All above discussions of the correlation paradox of the dissociation limit are based on the assumption that only the $1s$ shell orbitals of the two hydrogen nuclei are active, and that there is exactly one electron at each center at the dissociation limit.
\begin{figure}[htb]
    \centering
    \subfigure[Two-center dissociation]{\includegraphics[scale=0.13]{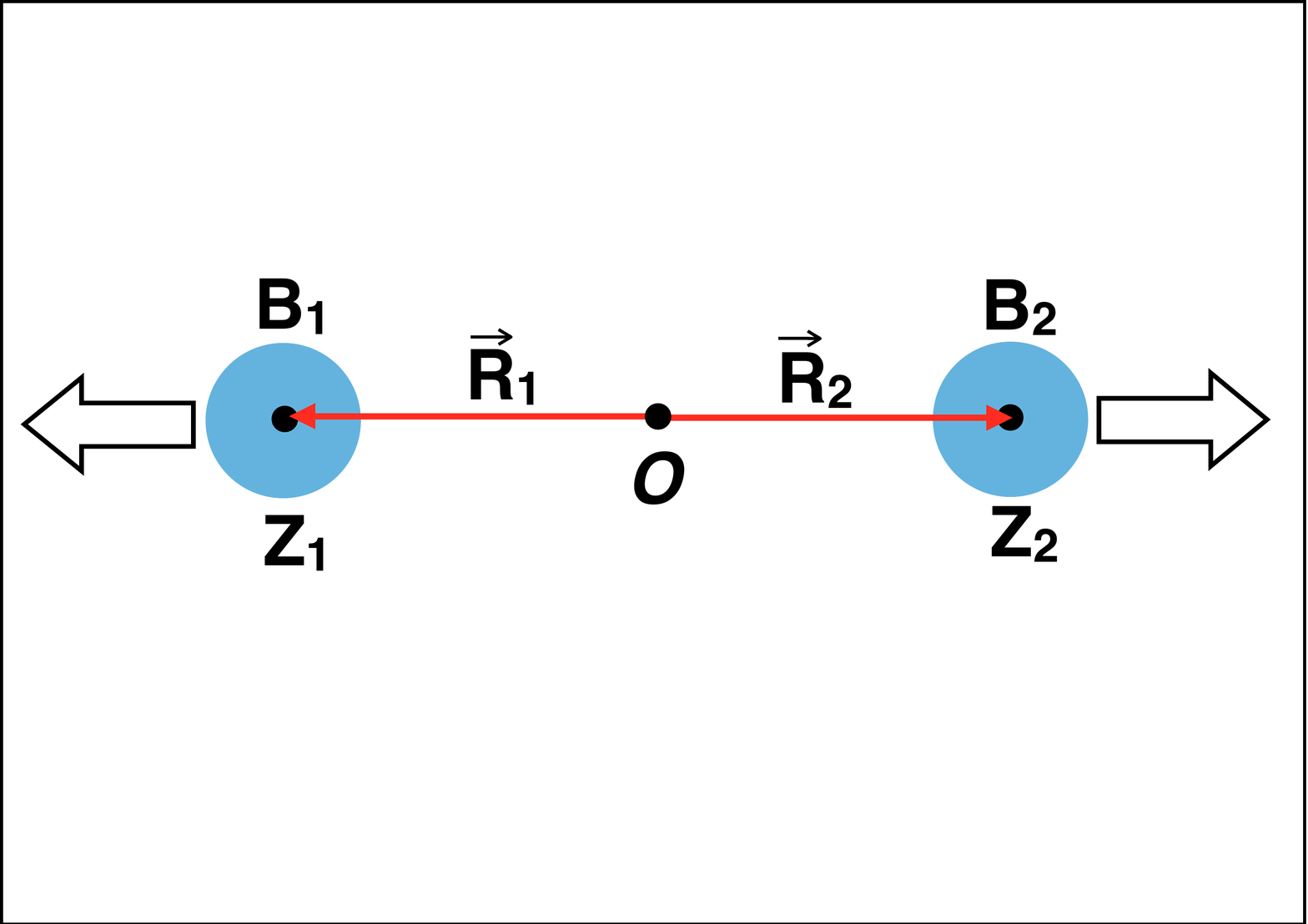}}
    \quad
    \subfigure[Five-center dissociation]{\includegraphics[scale=0.13]{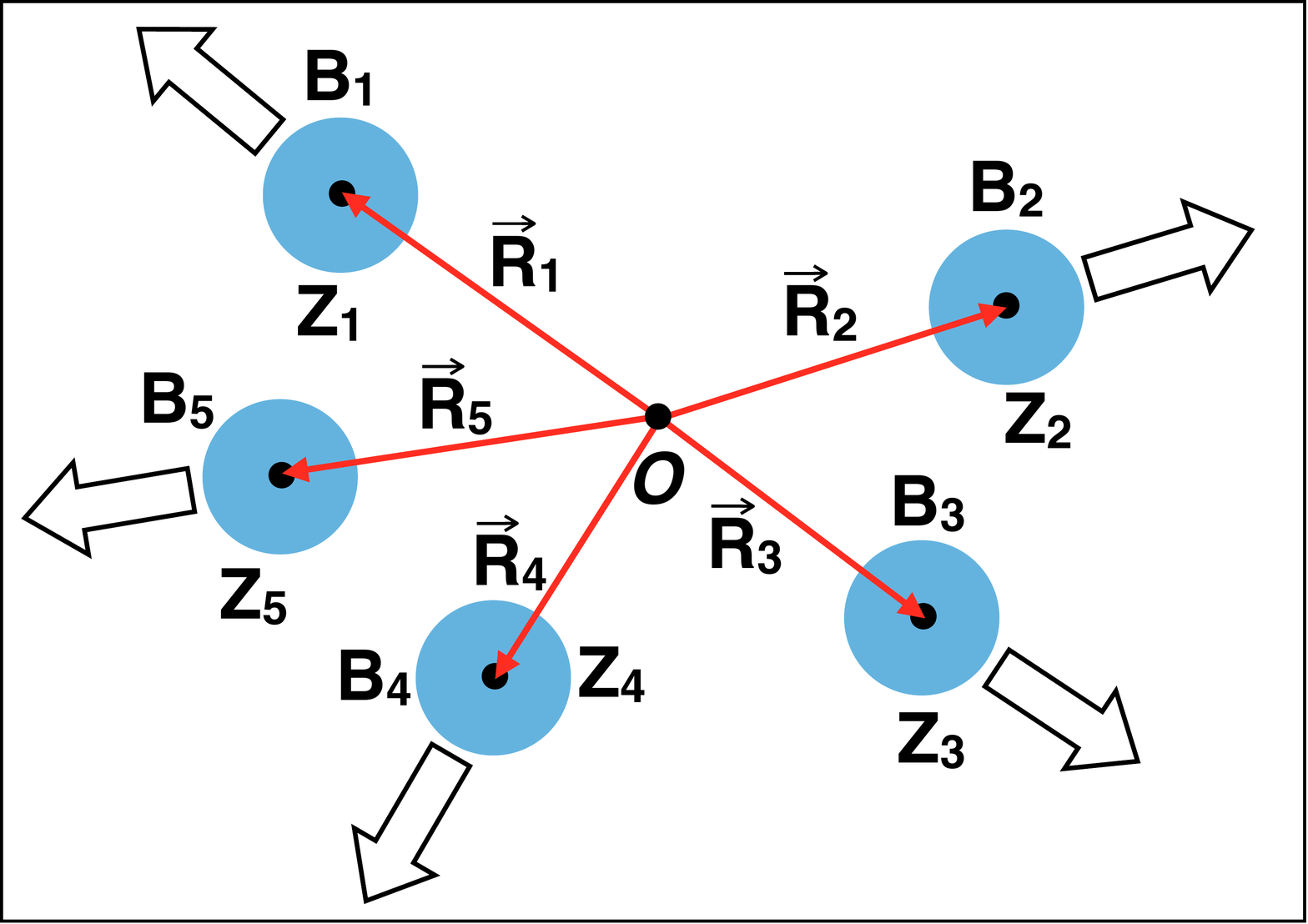}}
    \caption{Schematic illustration of dissociation in general: Various nuclei together with their local bases $B_i$ of atomic spin-orbitals centered at $\vec{R}_i$ are separated from each other (see text for more details).}
    \label{Dissociation}
\end{figure}
In the following we successively relax these assumptions to formulate a hierarchy of generalized correlation paradoxes in the dissociation limit.
In analogy to Sections \ref{sec:diss}, \ref{sec:resol}, we resolve those paradoxes by referring to thermal noise which will destroy in the dissociation limit the mode entanglement between different nuclear centers (and if applicable the quantum nonfreeness).

As illustrated in Figure \ref{Dissociation}, we consider a general molecular system with $\nu$ nuclear centers $Z_i$ at positions $\vec{R}_i$, $i=1,\ldots, \nu$. We then choose finite local basis sets $B_i = \{|\varphi^{(k)}_i\rangle\}_{k=1}^{d_{i}}$
of atomic spin-orbitals which are localized mainly around the respective center $Z_i$. The general dissociation limit can then be formally described, e.g., by the process $\Vec{R}_i \rightarrow \lambda \Vec{R}_i$ as $\lambda \rightarrow \infty$, where $\lambda$ is a scale parameter, e.g., $\lambda=1$ could correspond to the equilibrium geometry of the molecule. Actually, 
it is only crucial for the following considerations that the nuclei separate from each other more and more in the dissociation limit, i.e., $|\vec{R}_i-\vec{R}_j|\rightarrow \infty$ for all $i,j$. Scenarios in which two or more nuclei remain at finite separation distances are included as well and in that case we would merge them to form one joint (more complex) center.
Moreover, we denote the local one-particle Hilbert space at center $Z_i$ by $\mathcal{H}_i = \mathrm{Span}(B_i)$. The corresponding local Fock spaces $\mathcal{F}_i$ follow as the (direct) sums of different fixed particle number sectors $\wedge^N[\mathcal{H}_i]$ generated by $\mathcal{H}_i$,
\begin{equation}
    \mathcal{F}_i = \bigoplus_{N_i=0}^{\scriptsize{\mbox{dim}}(\mathcal{H}_i)} \wedge^{N_i}[\mathcal{H}_i]. \label{Fock}
\end{equation}
For each center $Z_i$ a natural basis $\overline{B}_i$ for its Fock space $\mathcal{F}_i$ is given by the family of configuration states Eq.~\eqref{config1st} which involve only spin-orbitals belonging to $B_i$. The corresponding local observables, i.e., Hermitian operators acting on  $\mathcal{F}_i$ form a local algebra, $\mathcal{A}_i$. Physically admissible operators obey number parity (or particle number) superselection rules, that is, when represented in $\overline{B}_i$, they are block diagonal with respect to the even and odd particle number sector (or all particle number sectors) and thus preserve the number parity (or particle number).

The Fock space $\mathcal{F}$ of the total system is given by the  tensor product of all local Fock spaces,
\begin{equation}\label{Focktotal}
    \mathcal{F} = \bigotimes_{i=1}^\nu \mathcal{F}_i.
\end{equation}
Since the molecular system has a fixed particle number $N$, we could restrict to the corresponding particle number sector of $\mathcal{F}$.
The electronic Hamiltonian $\hat{H}$ of the molecular system expressed in second quantization can be decomposed into local terms $\hat{H}_i$ and ``coupling'' terms $\hat{H}_{ij}$,
\begin{equation}\label{Hdecomp}
    \hat{H} = \sum_{i=1}^\nu \hat{H}_i + \sum_{1 \leq i < j \leq\nu} \hat{H}_{ij}.
\end{equation}
The local terms $\hat{H}_i$ involve only creation and annihilation operators referring to the spin-orbitals of center $Z_i$ while the coupling terms $\hat{H}_{ij}$ refer to two centers $Z_i,Z_j$. The latter ones describe the Coulomb pair interaction of electrons/nuclei at center $Z_i$ with those at center $Z_j$ and the hopping of the electrons between those centers (kinetic energy). Consequently, they decay in the dissociation limit, fulfilling
\begin{equation}\label{couplingdecay}
  \|\hat{H}_{ij}\| \leq \frac{q_{ij}}{|\vec{R}_i-\vec{R}_j|}
\end{equation}
with some appropriate constants $q_{ij}$. In contrast to those coupling terms $\hat{H}_{ij}$, the local terms $H_i$ are effectively independent of the dissociation limit (only their reference points $\vec{R}_i$ change).

After having formally introduced the general physical system, we can now present a hierarchy of generalized dissociation limits and their resolution on a \emph{qualitative} level:

\begin{enumerate}
\item \emph{Full Electron Separation}. In this scenario, we assume at the dissociation limit that each nuclear center will be occupied by exactly one electron. This is, only the one-particle sectors in the local Fock spaces $\mathcal{F}_i$ in Eq.~\eqref{Fock} are occupied. Prime examples of this situation are the dissociation of the hydrogen molecule $\mathrm{H}_2$, its isotopic variations $\mathrm{HD}$ and $\mathrm{D}_2$ and just any chain or ring of hydrogen atoms.

    Since the dissociation limit spatially separates all electrons, their interaction is also marginalized. Naively, one may therefore expect that the dissociated ground state would take the form of a configuration state,
    \begin{equation}\label{PsiIa}
    \ket{\Psi_0} = \ket{\phi_1}\wedge \ldots \wedge \ket{\phi_{\nu}},
    \end{equation}
    involving from each center $Z_i$ its local one-electron ground state $\ket{\phi_i}$. In fact, this is the case as long as the local ground states $\ket{\phi_i}$ are nondegenerate for all (or all except one) centers $Z_i$. Otherwise, by adding to each $\ket{\phi_i^{(m_i)}}$ a superindex reflecting its possible  degeneracies, the respective $N$-electron ground state will typically take the form of a coherent superposition
    \begin{equation}\label{PsiI}
    \ket{\Psi_0} = \sum_{m_1,\ldots,m_\nu} a_{m_1,\ldots,m_\nu} \,\ket{\phi_1^{(m_1)}}\wedge \ldots \wedge \ket{\phi_{\nu}^{(m_{\nu})}}.
    \end{equation}
    For centers with a unique, non-degenerate ground state the respective sum collapses to just one term, $m_i=1$.

    Since at the dissociation limit the couplings $\hat{H}_{ij}$ between any two centers vanish \eqref{couplingdecay}, all configurations involved in \eqref{PsiI} have the same energy in that limit. This implies, that the presence of any finite temperature $T$ would turn  the state \eqref{PsiI} into a classical mixture of its configuration states and in that sense resolve the correlation paradox. Since for each of those configuration states the involved one-particle states $\ket{\phi_i^{(m_i)}}\in \mathcal{H}_i$ belong to a definite center (i.e., they are not coherent superpositions of spin-orbitals of different centers), exactly the same will hold true for the mode entanglement and mode correlation between any two centers. 

    \item \emph{Fixed Local Particle Number}. We relax the restriction of having only one electron per center $Z_i$ at the dissociation limit. Yet, we still assume fixed local electron numbers $N_i$, where $\sum_{i=1}^\nu N_i = N$. This type of situations arises, e.g., when a molecule dissociates into two (or more) neutral identical atoms, e.g., $\mathrm{N}_2$ and $\mathrm{O}_2$. Since the $N_i$ electrons at each center are not getting separated there is no reason to expect that the dissociated $N$-electron state would  be a configuration state. Indeed, the non-vanishing interaction between the electrons at each center can give rise to finite correlations.

        Nonetheless, the vanishing of the coupling terms implies that the ground state problem decouples into those of the individual centers.  To be more specific, one just needs to determine the local $N_i$-electron ground state $\ket{\Phi_i}\in \wedge^{N_i}[\mathcal{H}_i]$ of $\hat{H}_i$ at each center $Z_i$. Naively, due to the vanishing of the coupling terms $\hat{H}_{ij}$ at the dissociation limit \eqref{couplingdecay} one may then expect a dissociated ground state of the form
        \begin{equation}\label{PsiIIa}
        \ket{\Psi_0} = \ket{\Phi_1}\wedge \ldots \wedge \ket{\Phi_\nu},
        \end{equation}
        which could be seen as a generalized configuration state. 
        In fact, this is the case as long as the local $N_i$-electron ground states $\ket{\Phi_i}$ are nondegenerate for all (or all except one) centers $Z_i$. Otherwise, by adding to each $\ket{\Phi_i^{(m_i)}}$ a superindex reflecting its possible degeneracies, the respective $N$-electron ground state will typically take the form of a coherent superposition
        \begin{equation}\label{PsiIIb}
        \ket{\Psi_0} = \sum_{m_1,\ldots,m_\nu} a_{m_1,\ldots,m_\nu} \,\ket{\Phi_1^{(m_1)}}\wedge \ldots \wedge \ket{\Phi_{\nu}^{(m_{\nu})}}.
        \end{equation}
        Since at the dissociation limit the couplings $\hat{H}_{ij}$ between any two centers vanish \eqref{couplingdecay}, all generalized configuration states $\ket{\Phi_1^{(m_1)}}\wedge \ldots \wedge \ket{\Phi_{\nu}^{(m_{\nu})}}$ involved in \eqref{PsiIIb} have the same energy in that limit. This implies that the presence of any finite temperature $T$ would turn the state \eqref{PsiIIb} into a classical mixture of those wedge products with equal weights and in that sense resolves this generalized correlation paradox. Again, since each element $\ket{\Phi_i^{(m_i)}}$ belongs in the mode/orbital picture to a local Fock space $\mathcal{F}_i$, exactly the same applies to the mode entanglement and mode correlation between any two centers.

    \item \label{generalparadox} \emph{Mixed Local Particle Numbers}. We relax the assumptions even further, and now allow for mixed local particle numbers. This may even include cases in which the total system is coupled to an environment and therefore may have an indefinite total electron number. Typical example for isolated systems are molecules with an excess or shortage of electrons, such as  $\mathrm{N}_2^+$: At the dissociation limit of $\mathrm{N}_2^+$ the total thirteen-electron state will (in the simplest case) be an equal superposition of two generalized configuration states \eqref{PsiIIa}, one with seven electrons on the left and six on the right, and one with seven electrons on the left and six on the right,
    \begin{equation}\label{PsiIIIa}
        \ket{\Psi_0} = \frac{1}{\sqrt{2}}\left[\ket{\Psi_6^{(L)}} \wedge \ket{\Psi_7^{(R)}} + \ket{\Psi_7^{(L)}} \wedge \ket{\Psi_6^{(R)}}\right].
    \end{equation}
    Since even the simplest possible state \eqref{PsiIIIa} assumed for $\mathrm{N}_2^+$ does not take the form of a generalized configuration state \eqref{PsiIIa}, we have to give up the particle picture (wedge product-based notation) and consider exclusively the mode/orbital picture which is based on second quantization.
    Just to reiterate, the mode reduced density operators are defined with respect to the tensor product structure \eqref{Focktotal}. For instance, the mode reduced density operator for the mode subsystem $B_i$ and $\mathcal{H}_i$, respectively, at center $Z_i$ is obtained by taking the partial trace of the total state $\rho$ with respect to all factors $\mathcal{F}_j$, $j\neq i$. This leads to a reduced density operator acting on the local space $\mathcal{F}_i$ which in general does not have a definite particle number anymore. Yet, as long as the total state $\rho$ has a fixed particle number, any mode reduced density operator is block-diagonal with respect to the different particle number sectors $\wedge^{N_i}[\mathcal{H}_i]$.

    In the mode/ortbial picture, the consequences of the decay \eqref{couplingdecay} of the coupling terms $\hat{H}_{ij}$ are obvious: Since at the dissociation limit, the spin-orbitals belonging to different centers $Z_i$ do not couple anymore, one may naively expect that the corresponding $Z_i$-mode reduced density operators $\rho_i$ would be uncorrelated, and the total state would take the form
    \begin{equation}\label{PsiIIIb}
        \rho = \rho_1 \otimes \cdots \otimes \rho_\nu.
    \end{equation}
    In case the system is isolated, each $\rho_i$ would be a pure state (with possibly indefinite particle number) on the local Fock space $\mathcal{F}_i$.
     As the example \eqref{PsiIIIa} already illustrates, this is not necessarily the case whenever the dissociated total system has a degenerate ground state space spanned by generalized configuration states \eqref{PsiIIa} with varying local particle numbers. Consequently, in case of a finite temperature $T$, the same happens as in the previous scenario of fixed local particle numbers: All contributing generalized configurations are classically mixed with equal weights (yet those configurations have no definite local particle numbers anymore). Consequently, the mode entanglement and mode correlation between any two centers vanishes in the dissociation limit, regardless of how small $T>0$ is.
\end{enumerate}

In the following we resolve those generalized correlation paradoxes also in a \emph{quantitative} way, at least in the mode/orbital picture. In particular, this will illustrate why and how a finite temperature in combination with the decaying behaviour of the coupling terms $\hat{H}_{ij}$ affects and eventually kills the mode entanglement and mode correlation in the dissociation limit.
First, it suffices to resolve the most general version of the correlation paradox, since the one for mixed local particle numbers contains
the other two scenarios as special cases. Second, for the sake of providing a resolution of those paradoxes, one can ignore superselection rules, as the entanglement and correlation \textit{without} superselection rules serve as upper bounds for the physically accessible entanglement and correlation\cite{bartlett2003entanglement}. To be more specific, we only need to show that this upper-bound goes to zero at the dissociation limit to resolve the paradox. Third, given a consistent correlation and entanglement measure, the total correlation is always greater or equal to its quantum part \eqref{CorrvsE}. Then, in order to show that the entanglement vanishes at the dissociation limit, it suffices to show the same for the total correlation (as quantified by the quantum mutual information \eqref{CorrMeasure} without superselection rules). Combining the above arguments, we can safely claim to resolve in the following various correlation paradoxes by proving that the quantum mutual information between any two centers becomes zero at the dissociation limit at any finite temperature.

In the important work \cite{wolf2008area}, a universal relation has been found between the correlation in multipartite quantum systems and the system's temperature and individual coupling terms. Here we give a demonstration of the underlying ideas by applying it first to the Hubbard dimer. Afterwards we repeat those steps in the context of general molecular system to resolve the correlation paradox in a quantitative way. The Hubbard dimer Hamiltonian \eqref{Hubbard} can be written as
\begin{equation}
    \hat{H} = \hat{H}_{L} + \hat{H}_{R}+ \hat{H}_{LR},
\end{equation}
where $\hat{H}_L$ and $\hat{H}_R$ are terms that act only on the left or right nucleus/site, and $\hat{H}_{LR}$ denotes the ``coupling'' between both mode subsystems, i.e., the hopping term. To refer more to the previous sections we have replaced here the indices of the nuclear centers according to $1 \mapsto L$ and $2 \mapsto R$.

The thermal equilibrium state $\rho$ \eqref{Gibbs} follows as the minimizer of the free energy
\begin{equation}\label{thermo1}
    F = E - T S.
\end{equation}
Here, $S$ denotes the von Neumann entropy of the total state $\rho$,  $E \equiv \langle \hat{H}\rangle_{\rho}\equiv \Tr[H\rho]$ and we
denote in the following the mode reduced density operators of $\rho$ for the left and right mode subsystem by $\rho_L$ and $\rho_R$, respectively. As a consequence of the characterization of $\rho$, the free energy of the state $\rho_L \otimes \rho_R$ is larger than that of $\rho$,
\begin{equation}\label{thermo2}
    \Tr[\hat{H}\rho] - TS(\rho) \leq \Tr[\hat{H}(\rho_L \otimes \rho_R)] - TS(\rho_L \otimes \rho_R).
\end{equation}
Equivalently, this can be stated as
\begin{equation}\label{thermo3}
    \Tr\left[\hat{H}\big(\rho-\rho_L \otimes \rho_R\big)\right]   \leq T \left[S(\rho) - S(\rho_L \otimes \rho_R)\right].
\end{equation}
Since the right-hand side of Eq.~\eqref{thermo3} is up to a prefactor $-T$ nothing else than the quantum mutual information of $\rho$
it follows that (for the sake of clarity we make explicit the chosen decomposition of the mode system into left and right)
\begin{equation}
    \begin{split}
        I_{\rho}(L:R) &\leq \frac{1}{T} \Tr[\hat{H}(\rho_L \otimes \rho_R - \rho)]
        \\
        & = \frac{1}{T} \Tr[\hat{H}_{LR} (\rho_L \otimes \rho_R - \rho)]
        \\
        & \leq \frac{2\|\hat{H}_{LR}\|_F}{T}. \label{wolflaw}
    \end{split}
\end{equation}
In the second line we have used $\Tr[\hat{H}_{i} (\rho_L \otimes \rho_R - \rho)]=0$ for $i=L/R$.
The remarkable relation \eqref{wolflaw} states that the quantum mutual information is bounded by the strength of the coupling/hopping term $\hat{H}_{LR}$, and decays to zero in the dissociation limit.

This whole illustration of the work \cite{wolf2008area} in the Hubbard dimer model and in particular \eqref{wolflaw} can also be generalized to a multivariate setting. For this one first defines the generalized quantum mutual information as\cite{watanabe1960information}
\begin{equation}
    I_\rho(1:2:\cdots : \nu) \equiv S(\rho || \rho_1 \otimes \rho_2 \otimes \cdots \otimes \rho_\nu).
\end{equation}
It quantifies the quantum information of the total state $\rho$ which is not yet contained in the single-center reduced density operators $\rho_1,\ldots,\rho_\nu$.  Since it is concerned with a decomposition of the total system into  several subsystems it will be particularly useful for our generalized dissociation limit beyond diatomic molecules.
Various steps of the derivation of \eqref{wolflaw} for the Hubbard dimer can be repeated in a similar fashion to any molecular system in its multi-nuclear dissociation limit as defined at the beginning of this section. To explain this, we first decompose the Hamiltonian of the molecular system according to \eqref{Hdecomp}.
Then, by recalling the characterization of the Gibbs state as the minimizer of \eqref{thermo1} and comparing its free energy to the one of the
product state $\rho_1 \otimes \rho_2 \otimes \cdots \otimes \rho_\nu$, we find
\begin{equation}
    F(\rho) \leq F(\rho_1 \otimes \rho_2 \otimes \cdots \otimes \rho_\nu).
\end{equation}
Plugging in the definition of the free energy and repeating the steps below \eqref{thermo2} immediately leads to the desired final result
\begin{equation} \label{generallaw}
    I_{\rho}(1:2:\cdots:\nu) \leq \frac{2}{T}  \sum_{1 \leq i < j \leq\nu} \|\hat{H}_{ij}\|_F.
\end{equation}
Relation \eqref{generallaw} in combination with the decay \eqref{couplingdecay} of the coupling terms $\hat{H}_{ij}$ implies that for any finite temperature $T>0$ the thermal state of the molecular system converges to the mode uncorrelated state $\rho_1 \otimes \rho_2 \otimes \cdots \otimes \rho_\nu$ in the dissociation limit. Hence, the correlation paradox is completely resolved in the mode/orbital picture:
The mode correlation between any two nuclear centers and thus also any correlation function (recall \eqref{CvsI}) vanishes in the dissociation limit for all molecular quantum systems under realistic experimental conditions. This is due to the presence of thermal noise, regardless of how close the temperature is to the absolute minimum of zero Kelvin.

\section{Summary and Conclusion}\label{sec:concl}
Particularly in dissociation limits, the interplay between electron interaction and geometry of a molecular system can lead to rather paradoxical situations. For instance for the hydrogen molecule, the exact ground state in the dissociation limit does not take the form of a single configuration state/Slater determinant despite the fact that increasing the separation distance between both nuclei marginalizes the electron-electron interaction. In our work we have formally introduced such correlation paradoxes in dissociation limits
and generalizations thereof to molecular systems with several electrons at each nuclear center. The aim was then to resolve them in a quantitative way based on concise quantum information theoretical tools.

This aim required a detailed introduction into the concept of entanglement and total correlation. Therefore, we have first reviewed foundational aspects of those concepts, worked out in quantum information theory for systems of distinguishable subsystems. Since mainly entanglement represents a fundamentally important resource for realizing exciting information theoretical tasks its operationally meaningful quantification in concrete systems is crucial. We have therefore illustrated how those concepts could be transferred to chemical systems of identical fermions. Since entanglement is relative in the sense that it always refers to a (typically non-unique) decomposition of the total system into subsystems there are different routes for identifying such a structure. By referring to second quantization a notion of so-called mode/orbital entanglement and correlation follows naturally while first quantization leads to the concept of (quantum) nonfreeness\cite{gottlieb2005new,gottlieb2007properties,Gottlieb14,Gottlieb15,schliemann2001quantum}. The latter quantifies the distance of quantum states to the closest configuration state/Slater determinant. This is also the reason why the nonfreeness provides a promising concise tool for quantifying the intrinsic computational complexity of ground state problems\cite{Pachos17free,Pachos18free,Pachos19free}. In that context, we have not only explained how to introduce respective measures of those concepts but have also shown how commonly ignored obstacles can be overcome. Examples of such obstacles are the number parity and particle number superselection rules. They represent fundamental physical limitations and ignoring them would lead to an overestimation of the extractable total correlation and entanglement in concrete quantum systems.

From a general point of view, we also believe that concise concepts from quantum information theory could play an even more prominent role in quantum-many body physics in the future:  They may help to identify and exploit all structural simplifications of realistic Hamiltonians to  develop more effective and accurate solutions of the ground state problem: On the one hand, the electrons interact only by two-body forces which would imply a significant simplification in the \emph{particle picture}. Indeed, the variational ground state problem could in principle be solved in terms of the two-\emph{particle} reduced density matrix\cite{Col63,Percus64,Mazz12,Mazz16}. On the other hand, the pair interaction decays in spatial space which implies a decay of correlation functions in the \emph{mode/orbital picture}. A remarkable result along those lines is that particle correlation equals in some sense mode/orbital correlation minimized with respect to all mode/orbital bases\cite{Gigena15}.

Equipped with those tools, we have resolved the correlation paradoxes in the dissociation limits of molecular systems in a quantitative way: We have proven that thermal noise due to temperature will destroy the mode entanglement beyond a critical separation distance $r_{\mathrm{crit}}^{(m)}$($T$) and the total mode correlation at the dissociation limit entirely. This means that all correlation functions referring to different nuclear centers vanish in the dissociation limit, provided the temperature is finite. A study in the particle picture, has confirmed that thermal noise turns coherent superpositions of (quasi)degenerate configuration states into classical mixtures of them. As a matter of fact, the more general result \eqref{generallaw} emphasizes that any form of perturbation of the system would have the same effect in the dissociation limit as thermal noise. Hence, from a practical point of view, our findings emphasize that neither finite mode entanglement nor finite quantum nonfreeness can ever be observed in the dissociation limit in a laboratory. This also rationalizes and clarifies the perception that the ``correlation'' of the dissociated ground state vanishes.

Based on the important work \cite{wolf2008area}, we could explain why and how the presence of thermal noise in combination with the decay of certain interaction terms in the dissociation limit leads to a vanishing of several inter-nuclei correlation functions. Since those resemble exactly the static correlation of the system, our quantitative analysis suggests a new characterization of static and dynamic correlations: The more robust the total correlations are with respect to thermal noise the stronger the dynamic correlations $C_{dyn}$ in the system. To be more specific, one may even take the ratio of the left- and right-hand side of Eq.~\eqref{generallaw} (or any improved version of it) as a measure $C_{dyn}$ for dynamic correlation (and $C_{stat} \equiv 1-C_{dyn}$ for the static ones). Indeed, according to the right-hand side the presence of thermal noise destroys a part of the total correlation. If the actual reduction (left-side) is, however, lower than it has to be, the system necessarily had some static correlations (since those are apparently much more unstable with respect to thermal noise than dynamic ones). Or conversely, if for some temperature $T$ the inequality \eqref{generallaw} is saturated, it means that the effect of thermal noise was as minimal as it could have been and the ground state could not have contained (much) static correlation.


\section{Acknowledgment}
We thank N.\hspace{0.5mm}Friis, B.\hspace{0.5mm}Morris, S.\hspace{0.5mm}Pittalis, S.\hspace{0.5mm}Szalay, V.\hspace{0.5mm}Vedral, \mbox{B.\hspace{0.5mm}Yadin} and particularly Z.\hspace{0.5mm}Zimboras for helpful discussions. CS acknowledges financial support from the Deutsche Forschungsgemeinschaft (Grant SCHI 1476/1-1) and the UK Engineering and Physical Sciences Research Council (Grant EP/P007155/1).

\appendix

\section{Spectrum of Hubbard Dimer}\label{spectrum}
The Hubbard dimer model contains four spin-orbitals $\{\ket{L\!\uparrow}, \ket{L\!\downarrow}, \ket{R\!\uparrow}, \ket{R\!\downarrow}\}$ which span together the underlying one-particle Hilbert space $\mathcal{H}_1$. The total Fock space is given as the (direct) sum of various particle number sectors $\mathcal{H}_N=\wedge^N[\mathcal{H}_1]$,
\begin{equation}
    \mathcal{F} = \bigoplus_{N=0}^4 \mathcal{H}_N
\end{equation}
Since we consider the Hubbard dimer as an effective model for the hydrogen molecule in the dissociation limit, we restrict ourselves to the \(N=2\) sector $\mathcal{H}_2$ which has dimension $\binom{4}{2}=6$. We can divide $\mathcal{H}_2$ into spin sectors with magnetization \(M = -1, 0, 1\).
\begin{enumerate}
    \item \(M = \pm1\). Only one possible state in each sector:
    \begin{equation}
        |\Psi_{\uparrow/\downarrow}\rangle  = f^\dagger_{L\uparrow/\downarrow} f^\dagger_{R\uparrow/\downarrow} |0\rangle,
    \end{equation}
    which is therefore also an eigenstates of the Hamiltonian \eqref{Hubbard}. Its  energy is $0$ since no hopping is allowed and harboring two electrons with the same spin is forbidden by the Pauli exclusion principle.

    \item \(M = 0\). A basis of this sector contains four states, which can be grouped into different reflection parity sectors, denoted by $p=\pm$:
    \begin{eqnarray}
        |1_\pm \rangle = \frac{1}{\sqrt{2}}(f^\dagger_{L\uparrow}f^\dagger_{R \downarrow} \mp f^\dagger_{L\downarrow}f^\dagger_{R\uparrow})|0\rangle, \nonumber
        \\
        |2_\pm \rangle = \frac{1}{\sqrt{2}}(f^\dagger_{L\uparrow}f^\dagger_{L \downarrow} \pm f^\dagger_{R\uparrow}f^\dagger_{R\downarrow})|0\rangle.
    \end{eqnarray}
    The state $\ket{1_-}$ belongs to the triplet ($S=1$) while the other three are singlets ($S=0$).
    Energy eigenstates can be found via exact diagonalization:
    \begin{eqnarray}
    E_0 &=& \frac{U}{2} - W, \quad |\Psi_0\rangle = a |1_+\rangle + b |2_+\rangle, \nonumber
    \\
    E_1 &=& 0, \hspace{1.3cm} |\Psi_1\rangle = |1_-\rangle, \nonumber
    \\
    E_2 &=& 0, \hspace{1.3cm} |\Psi_2\rangle = |2_-\rangle, \nonumber
    \\
    E_3 &=& \frac{U}{2} + W, \quad |\Psi_3\rangle = c |1_+\rangle + d |2_+\rangle,
    \end{eqnarray}
    where
    \begin{equation}
     W = \sqrt{\frac{U^2}{4}+4t^2}
    \end{equation}
    and
    \begin{eqnarray}
    a = \sqrt{\frac{W+\frac{U}{2}}{2W}}, \quad b = \frac{2t}{\sqrt{2W\left(W+\frac{U}{2}\right)}}, \nonumber
    \\
    c = -\sqrt{\frac{W-\frac{U}{2}}{2W}}, \quad d = \frac{2t}{\sqrt{2W\left(W-\frac{U}{2}\right)}}.
    \end{eqnarray}

\end{enumerate}

\section{Quantum Mutual Information as Correlation Measure}\label{app:MI}

Given full access to local operators, a state $\sigma$ is uncorrelated if and only if it factorizes according to $\sigma = \sigma_A \otimes \sigma_B$. The quantum relative entropy of a general state $\rho$ relative to $\sigma_A \otimes \sigma_B$ then follows as
\begin{eqnarray}
\lefteqn{S(\rho||\sigma_A \otimes \sigma_B)} && \nonumber \\
       &=& \Tr[\rho\log(\rho)] - \Tr[\rho\log(\sigma_A \otimes \sigma_B)] \nonumber
        \\
        &=& \Tr[\rho\log(\rho)] - \Tr[\rho\log((\sigma_A \otimes \mathbbm{1})(\mathbbm{1}\otimes \sigma_B))] \nonumber
        \\
        &=& \Tr[\rho\log(\rho)] - \Tr[\rho_A \log(\sigma_A)] - \Tr[\rho_B \log(\sigma_B)] \nonumber
        \\
        & =& S(\rho||\rho_A\otimes \rho_B) + S(\rho_A||\sigma_A) + S(\rho_B || \sigma_B) \nonumber
        \\
        & \geq &S(\rho||\rho_A\otimes \rho_B) = I(\rho).
\end{eqnarray}
The minimum of $ S(\rho||\sigma)$ with respect to all $\sigma=\sigma_A \otimes \sigma_B$
is obtained for $\sigma_A = \rho_A$ and $\sigma_B = \rho_B$. This follows form the previous derivation together with the non-negativity of the relative entropy. Hence the geometric correlation measure based on the quantum relative entropy \eqref{CorrMeasure} coincides with the quantum mutual information.

\section{$r_{\mathrm{crit}}^{(m/p)}(T)$ Divergence}\label{divergence}

In Figure \ref{fig:ME_sudden} and Figure \ref{fig:SuddenDeath} we presented the curve $r_{\mathrm{crit}}^{(m/p)}(T)$ above which the mode entanglement and the quantum nonfreeness vanished. In particular, we observed a diverging behavior when $T$ approaches zero. In this section we will determine the leading order of these divergences.

\textit{1. Mode/orbital Picture.} When $T$ is small, only the ground state and the first excitation level is activated. The local particle number superselected Gibbs state $\mathcal{T}_G(\rho) \equiv \tilde{\rho}$, whose entanglement gives the physical entanglement of the original Gibbs state under superselection rule, can be written as a sum of a separable state and a four-dimensional matrix which can be represented as
\begin{equation}
\begin{split}
    \tilde{\rho}|_{M_1} = \begin{pmatrix}
    e^{- \frac{\Delta E}{ T}} & 0 & 0 & 0
    \\
    0 & A & B & 0
    \\
    0 & B & A & 0
    \\
    0 & 0 & 0 & e^{- \frac{\Delta E}{T}}
    \end{pmatrix} \label{app:restricted}
\end{split}
\end{equation}
referring to the ordered basis states $f^\dagger_{L\uparrow} f^\dagger_{R\uparrow}|0\rangle$, $f^\dagger_{L\uparrow} f^\dagger_{R\downarrow}\ket{0}$, $f^\dagger_{L\downarrow} f^\dagger_{R\uparrow}\ket{0}$, $f^\dagger_{L\downarrow} f^\dagger_{R\downarrow}\ket{0}$ whose span is denoted by $M_1$. Here, we introduced
\begin{equation}
    A = \frac{1}{2}|a|^2 + \frac{1}{2} e^{- \frac{\Delta E}{ T}}, \quad B = -\frac{1}{2}|a|^2 + \frac{1}{2} e^{- \frac{\Delta E}{T}},
\end{equation}
and $a$ is as defined in Appendix \ref{spectrum} and $\Delta E = E_1 - E_0$. It is clear that $\tilde{\rho}$ is separable if and only if the expression in Eq.~\eqref{app:restricted} is separable. By the Peres-Horodecki criterion, Eq.~\eqref{app:restricted} is separable if and only if it has positive partial transpose. Then given a small temperature $T$, $r_{\mathrm{crit}}^{(m)}$ is the inter-nuclei distance such that the partial transpose of Eq.~\eqref{app:restricted} becomes rank deficient. That is, $r = r_{\mathrm{crit}}^{(m)}$ when
\begin{equation}
   A^2 (e^{- \frac{2\Delta E}{ T}} - B^2) = 0. \label{criticalcond}
\end{equation}
Since the factor $A^2$ cannot vanish we just need to solve $e^{- \frac{\Delta E}{ T}} = B$, leading to
\begin{equation}
3e^{- \frac{\Delta E}{T}}=a^2=\frac{1}{2} \left(\frac{1}{\sqrt{16 t^2+1}}+1\right).
\end{equation}
Resorting to the software Mathematica and recalling $t\equiv e^{-r}$ then yields the final result
\begin{equation}
    r_{\mathrm{crit}}^{(m)} = - \frac{1}{2} \log(T) + c_0 +c_1 T+\mathcal{O}(T^2), \quad T \rightarrow 0,
\end{equation}
where $c_0\equiv \log(2) - \frac{1}{2} \log(\log(3))$, $c_1\equiv -\frac{1}{2}(1+\log (3))$ are constants.

\textit{2. Particle Picture.} We recall the ``separability'' criterion from Eq.~\eqref{PartEnt}. We can similarly make the approximation by neglecting higher excitation contributions in the Gibbs state $\rho(T,r)$ as $T$ is small enough compare to $U\equiv 1$. In that case, the matrix $K(\rho)$ defined in Eq.~\eqref{C_SL} follows as
\begin{equation}
    K(\rho) = \begin{pmatrix}
    (b^2-a^2)p^2 & 0 & 0 & 0
    \\
    0 & q^2 & 0 & 0
    \\
    0 & 0 & 0 & -q^2
    \\
    0 & 0 & -q^2 & 0
    \end{pmatrix},
\end{equation}
where $a,b$ are defined as in Appendix \ref{spectrum}, and
\begin{equation}
    p = \sqrt{\frac{1}{1+3e^{- \frac{\Delta E}{ T}}}}, \quad q = \sqrt{\frac{e^{- \frac{\Delta E}{ T}}}{1 + 3 e^{- \frac{\Delta E}{ T}}}},
\end{equation}
with $\Delta E = E_1 - E_0$ as before. Recall the quantum nonfreeness $E^{(p)}(\rho)$ \eqref{PartEnt} is calculated as the largest absolute eigenvalue of $K(\rho)$ minus the absolute values of the rest. It is not hard to see that the absolute eigenvalues of $C(\rho)$ are $|a^2-b^2|p^2$ and $q^2$ where the latter is triply degenerate. At the point $r = r_{\mathrm{crit}}^{(p)}$ at which $E^{(p)}(\rho)$ reaches the value zero, the largest absolute eigenvalue can only be $|a^2-b^2|p^2$, and it must satisfies
\begin{equation}
    |a^2-b^2|p^2 = 3q^2.
\end{equation}
This leads to (using again the software Mathematica)
\begin{equation}
    r_{\mathrm{crit}}^{(p)} = - \frac{1}{2} \log(T) + d_0 +d_1 T+\mathcal{O}(T^2), \quad T \rightarrow 0,
\end{equation}
where $d_0\equiv \log(2) - \frac{1}{2} \log(\log(3))$, $d_1\equiv -\frac{1}{2}(2+\log (3))$ are constants.

\bibliography{references}

\end{document}